\RequirePackage{fix-cm}
\documentclass[smallextended]{svjour3}
\smartqed
\usepackage{appendix}
\usepackage{amsmath,amssymb}
\usepackage{graphicx}
\usepackage{lineno}
\usepackage{array}
\usepackage{longtable}
\usepackage{natbib}
\usepackage{xcolor}
\setcitestyle{aysep={}}

\newcommand*\patchAmsMathEnvironmentForLineno[1]{%
\expandafter\let\csname old#1\expandafter\endcsname\csname #1\endcsname
\expandafter\let\csname oldend#1\expandafter\endcsname\csname end#1\endcsname
\renewenvironment{#1}%
{\linenomath\csname old#1\endcsname}%
{\csname oldend#1\endcsname\endlinenomath}}%
\newcommand*\patchBothAmsMathEnvironmentsForLineno[1]{%
\patchAmsMathEnvironmentForLineno{#1}%
\patchAmsMathEnvironmentForLineno{#1*}}%
\AtBeginDocument{%
\patchBothAmsMathEnvironmentsForLineno{equation}%
\patchBothAmsMathEnvironmentsForLineno{align}%
\patchBothAmsMathEnvironmentsForLineno{flalign}%
\patchBothAmsMathEnvironmentsForLineno{alignat}%
\patchBothAmsMathEnvironmentsForLineno{gather}%
\patchBothAmsMathEnvironmentsForLineno{multline}%
}

\begin{document}

\title{Local Wind Regime Induced by Giant Linear Dunes: Comparison of ERA5-Land Reanalysis with Surface Measurements}

\author{Cyril Gadal \and Pauline Delorme \and Cl\'ement Narteau \and Giles F.S. Wiggs \and Matthew Baddock \and Joanna M. Nield \and Philippe Claudin}

\institute{
C. Gadal \at
Institut de M\'ecanique des Fluides de Toulouse, Universit\'e de Toulouse Paul Sabatier, CNRS, Toulouse INP-ENSEEIHT, Toulouse, France. \email{cyril.gadal@imft.fr}
\and
P. Delorme \at
Energy and Environment Institute, University of Hull, Hull, UK.
\at
School of Geography and Environmental Science, University of Southampton, Southampton, UK.
\and
C. Narteau \at
Université Paris Cité, Institut de physique du globe de Paris, CNRS, Paris, France
\and
Giles F.S. Wiggs \at
School of Geography and the Environment, University of Oxford, Oxford, UK.
\and
M.C. Baddock \at
Geography and Environment, Loughborough University, Loughborough, UK.
\and
J.M. Nield \at
School of Geography and Environmental Science, University of Southampton, Southampton, UK.
\and
P. Claudin \at
Physique et M\'ecanique des Milieux H\'et\'erog\`enes, CNRS, ESPCI Paris, PSL Research University, Universit\'e de Paris, Sorbonne Universit\'e, Paris, France.
}

\date{Received: DD Month YEAR / Accepted: DD Month YEAR}

\maketitle

{\vspace{-.5cm}\color{blue}\small  An edited version of this paper was published by Springer Nature: Gadal, C., Delorme, P., Narteau, C. et al. Local Wind Regime Induced by Giant Linear Dunes: Comparison of ERA5-Land Reanalysis with Surface Measurements. Boundary-Layer Meteorol (2022). https://doi.org/10.1007/s10546-022-00733-6 \vspace{.25cm}}.

\begin{abstract}
Emergence and growth of sand dunes results from the dynamic interaction between topography, wind flow and sediment transport. While feedbacks between these variables are well studied at the scale of a single and relatively small dune, the average effect of a periodic large-scale dune pattern on atmospheric flows remains poorly constrained, due to a pressing lack of data in major sand seas. Here, we compare local measurements of surface winds to the predictions of the ERA5-Land climate reanalysis at four locations in Namibia, both within and outside the giant linear dune field of the Namib Sand Sea. In the desert plains to the north of the sand sea, observations and predictions agree well. This is also the case in the interdune areas of the sand sea during the day. During the night, however, an additional wind component aligned with the giant dune orientation is measured, in contrast to the easterly wind predicted by the ERA5-Land reanalysis.
For the given dune orientation and measured wind regime, we link
the observed wind deviation (over 50\textdegree) to the daily cycle of the turbulent
atmospheric boundary layer. During the night, a shallow boundary layer
induces a flow confinement above the giant dunes, resulting in large
flow deviations, especially for the slower easterly winds. During the
day, the feedback of the giant dunes on the atmospheric flow is much weaker
due to the thicker boundary layer and higher wind speeds. Finally, we
propose that the confinement mechanism and the associated wind
deflections induced by giant dunes could explain the development of smaller-scale secondary
dunes, which elongate obliquely in the interdune areas of the primary dune
pattern.

\end{abstract}

\newpage

\section{Introduction}

The description of turbulent flows over complex topography is relevant for a large variety of different environmental systems \citep{Sherman1978, Walmsley1982, baines1995, Wood2000, Venditti2013, Finnigan2020}. For example, the flow over hills is of primary interest for wind power, meteorological and air pollution phenomena \citep{Taylor1987}. The properties of these flows are also key to the understanding of geophysical phenomena, including the formation of wind-driven waves on the ocean surface \citep{Sullivan2010}, dissolution bedforms \citep{Claudin2017, Guerin2020}, or sedimentary ripples and dunes \citep{Bagnold1941, Charru2013, Courrech2015}. Importantly, the troposphere presents a vertical structure, with a lower convective boundary layer, of typical kilometer-scale thickness, capped by a stably stratified region \citep{Stull1988}. The largest topographic obstacles, such as mountains, can therefore interact with this upper region and lead to internal wave generation or significant wind disturbances, such as lee-side downslope winds \citep{Durran1990}.

Compared to hills and mountains, aeolian sand dunes offer idealized elevation profiles for the study of atmospheric turbulent flow over topographies, due to their smooth shape, free of canopies. Besides, dunes provide a rather wide range of scales, from decameters to kilometers, and very often come in a fairly regular pattern, which further simplifies the flow structure analysis. Past studies have highlighted two important topographic feedbacks on the wind flow close to the dune/hill surface.
First is the effect on wind speed, with documented flow acceleration on upwind slopes \citep{Weaver2011} and deceleration on downwind slopes \citep{Baddock2007}, where the speed-up factor is essentially proportional to the obstacle aspect ratio \citep{Jackson1975}. Under multidirectional wind regimes with frequent wind reversals, this speed-up effect induces large differences in the amplitude and orientation of the resultant sediment transport between flat sand beds and the dune crests \citep{Zhang2014, Rozier2019, Gao2021}. In addition, the position of maximum velocity is typically shifted upwind of the obstacle crest \citep{Jackson1975, Claudin2013}. This behaviour has been theoretically predicted by means of asymptotic analysis of a neutrally stratified boundary-layer flow over an obstacle of vanishing aspect ratio \citep{Jackson1975, Mason1979, Sykes1980, Hunt1988, Belcher1998, Kroy2002}. Experiments in flumes \citep{Zilker1977, Zilker1979, Frederick1988, Poggi2007, Bristow2022}, in wind tunnels \citep{Gong1989, Finnigan1990, Gong1996} and in field conditions at all scales \citep{Taylor1987a, Claudin2013, Fernando2019, Lu2021}, have also documented this effect. Interestingly, a similar behaviour exists for the pressure perturbation, but with a slight downwind shift for the pressure minimum \citep{Claudin2021}.

The second effect, much less studied, is the flow deflection that occurs when the incident wind direction is not perpendicular to the ridge crest. While predicted to be small (less than $10^{\circ}$) in the linear regime valid for shallow topography \citep{Gadal2019}, significant flow steering has been reported in the field on the downwind side of steep enough obstacles, such as well-developed sand dunes \citep{Tsoar1983, Sweet1990, Walker2002, Smith2017} and in particular coastal foredunes \citep[e.g.][]{Hunter1983, rasmussen1989, Walker2006, Walker2009, Hesp2015, Walker2017, deWinter2020}, mountain ranges \citep{Kim2000, Lewis2008, Fernando2019}, and valley topographies \citep{Wiggs2002, Garvey2005}.

Wind measurements over sand dunes have been mainly performed over small bedforms, typically a few meters high (corresponding to several tens of meters long) \citep[e.g.][]{Mulligan1988, Hesp1989, Lancaster1996, MckennaNeuman1997, Sauermann2003, Andreotti2002, Walker2002, Weaver2011}. For practical reasons, fewer studies have performed similar measurements on giant dunes \citep{Havholm1988}, with kilometer-scale wavelengths and heights of tens of meters. However, such large dunes provide a choice configuration for the study of turbulent flows over a complex topography. First, one expects larger wind disturbances for larger obstacles. Secondly, their large size can make them interact with the vertical structure of the atmosphere \citep{Andreotti2009}. Third, they usually form large patterns in sand seas and thus behave as rather clean periodic perturbations, in contrast with isolated dunes. Finally, because the morphodynamics of aeolian bedforms is strongly dependent on the local wind regime \citep{Livingstone2019}, one can expect to see the consequences of windflow disturbance by large dunes on neighbouring small dunes \citep{Brookfield1977, Ewing2006}. A similar effect is observed on the properties of impact ripple patterns due to the presence of dunes \citep{Howard1977, Hood2021}.

Atmospheric flows have been much studied at the desert-scale with climate reanalyses based on global atmospheric models \citep{Blumberg1996, Livingstone2010, Ashkenazy2012, Jolivet2021, Hu2021, gunn2021}, such as ERA-40, ERA-Interim or ERA5 \citep{Uppala2005, Dee2011, Hersbach2020}. However, the spatial resolution of these reanalyses (tens of kilometers) implies average quantities that do not resolve the smaller scales of interest, which range from individual dunes to small mountains \citep{Livingstone2010}. Recently, the release of ERA5-Land has partly resolved this limitation by providing up to 70 years of hourly wind predictions at a $9$~km spatial resolution \citep{munoz2021}. However, its validity remains to be studied, especially in remote desert areas where assimilation of measured data is very low.

In this work, we compare local wind speeds and directions measured by meteorological stations at four different locations inside and north of the giant linear dune field of the Namib sand sea to the regional predictions of the ERA5-Land climate reanalysis. Where the meteorological stations are surrounded by a relatively flat environment, we show that local measurements and regional predictions agree well. The agreement is also good in the interdune areas of the sand sea, except for some weak winds blowing at night, which exhibit an additional component aligned with the giant dune orientation. These winds are not predicted by the ERA5-Land reanalysis (Sect.~\ref{sec:sec2}). Further, we are able to link the magnitude of these differences to the circadian cycle of the atmospheric boundary layer (Sect.~\ref{sec:sec3}). Finally, we draw implications for the wind disturbances on smaller-scale dunes (Sect.~\ref{sec:sec4}), suggesting a possible origin for crossing dunes, a distinctive secondary dune form observed in the Namib and other sand seas.

\begin{figure}[t]
  \centering
  \includegraphics[scale=1]{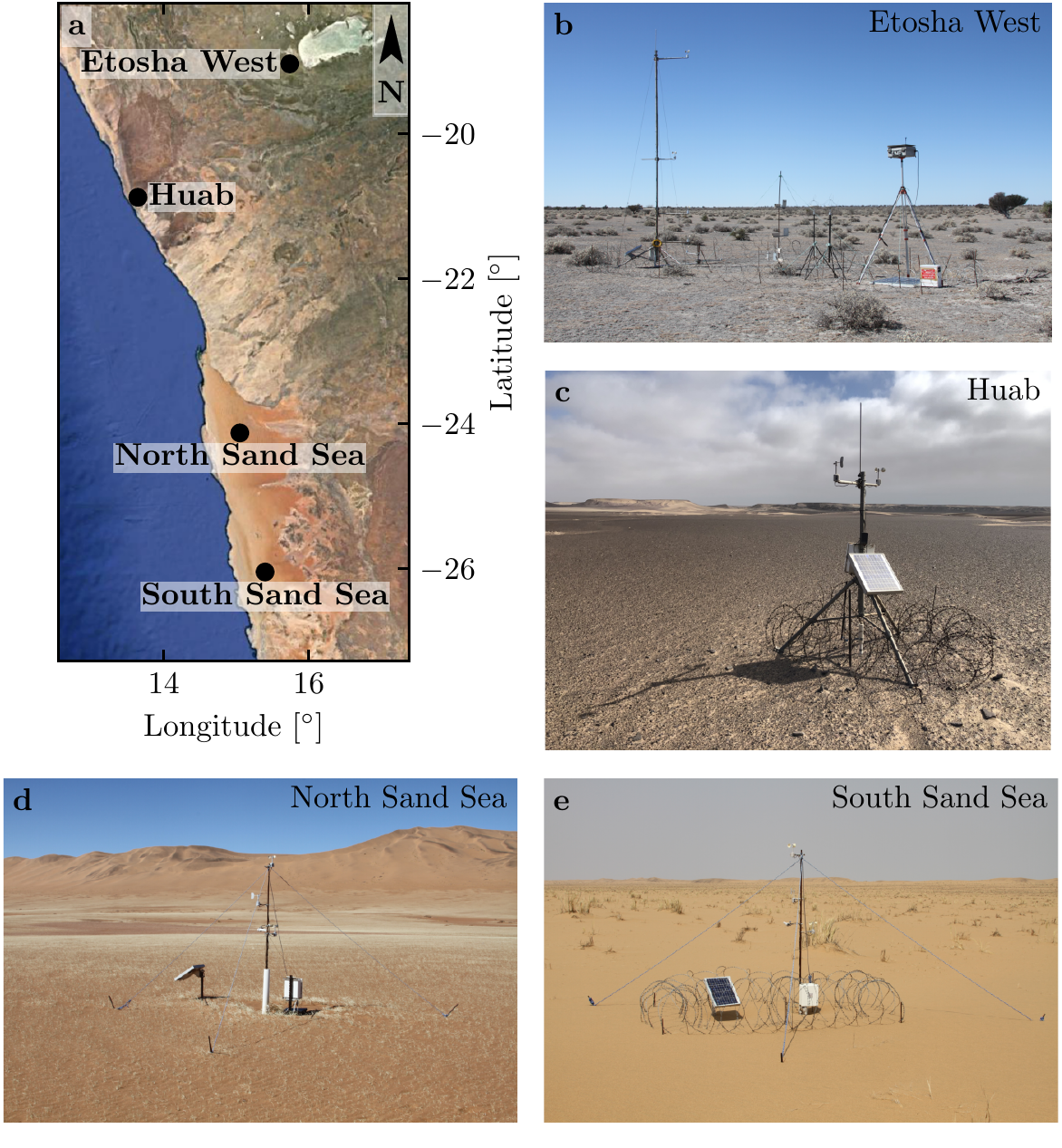}
  \caption{Studied field sites. \textbf{a} Location of the different sites in Namibia. \textbf{b-e}: Photographs of the meteorological stations.}
  \label{Fig1}
\end{figure}

\begin{figure}[t]
\centering
\includegraphics[scale=1]{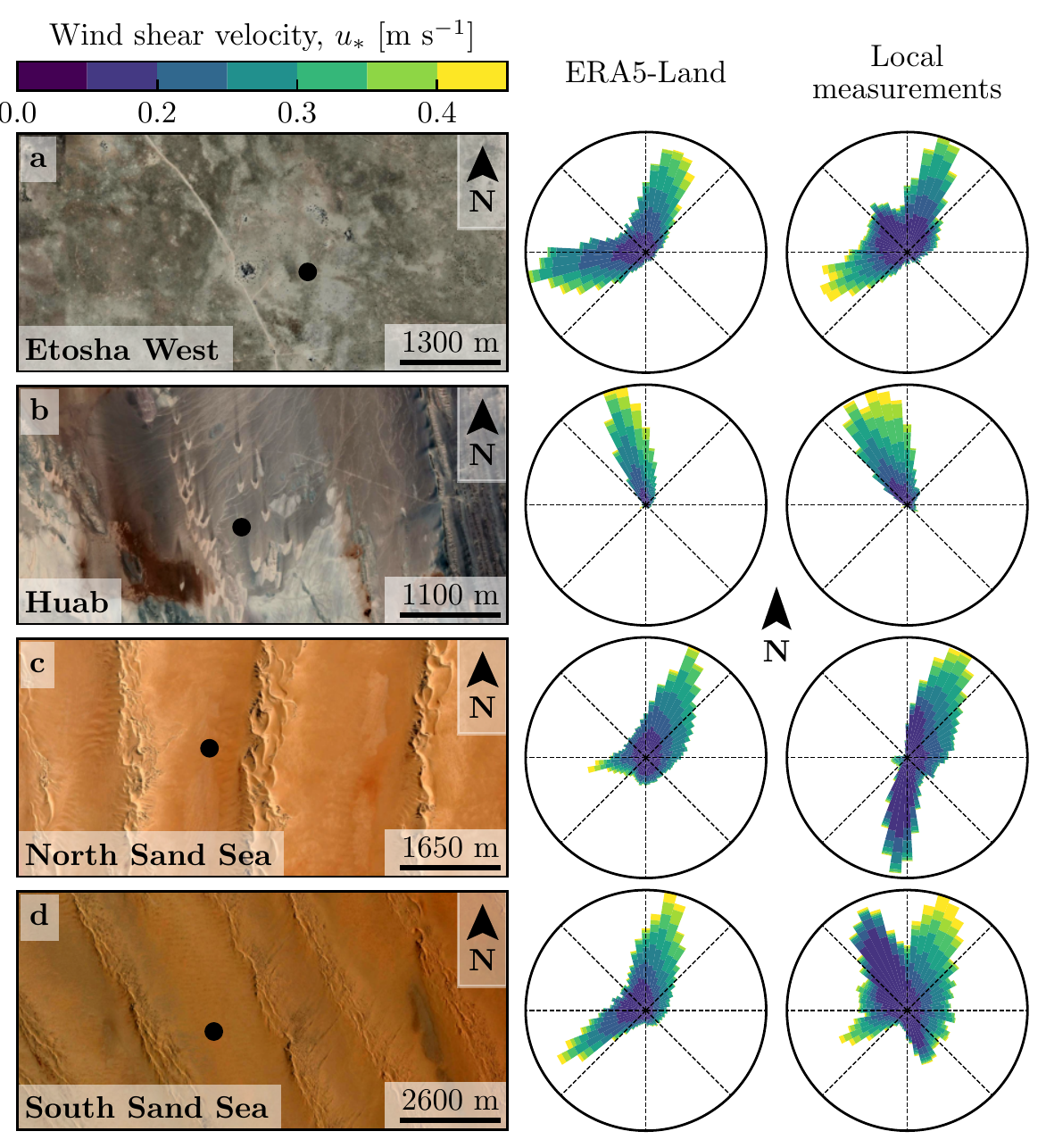}
\caption{Wind data used in this study. Satellite images of the different environments (Google-Earth, Maxar Technologies, CNES/Airbus) are shown on the left. The black dots show the location of the wind measurements stations. On the right of the photos, the corresponding wind roses representing the data from the ERA5-Land climate reanalysis and the local wind stations are displayed. Note: the graphical convention for the wind roses is that the bars show the direction towards which the wind blows (see color bar for velocity scale).}
\label{Fig2}
\end{figure}

\section{Wind Regimes Across The Namib Sand Sea}
\label{sec:sec2}
We measured the wind regime at four different locations in Namibia, representative of various arid environments across the Namib desert (Figs.~\ref{Fig1}~and~\ref{Fig2}). The Etosha West station was located at the Adamax waterhole to the west of Etosha Pan in northern Namibia, in a sparsely vegetated area. The Huab station was near the coast on a hyper-arid flat gravel plain lying north of the ephemeral Huab river. Here, barchan dunes up to a few meters in height develop from the sediment blowing out of the river valley \citep{Nield2017, Hesp1998}. These two stations were both located in relatively flat environments. In contrast, the North Sand Sea and South Sand Sea stations were located in the interdunes between linear dunes with kilometer-scale wavelengths, hectometer-scale heights and superimposed patterns. In this section, we describe and compare winds from local measurements and climate reanalysis predictions.

\subsection{Wind and Elevation Data}
At each meteorological station (Fig.~\ref{Fig1}), wind speed and direction were sampled every 10 minutes using cup anemometers (Vector Instruments A100-LK) and wind vanes (Vector Instruments W200-P) at a single height, which was between 2~m and 3~m depending on the station. The available period of measurements at each station ranged from 1 to 5 discontinuous years distributed between 2012 and 2020 (Online Resource Fig.~\ref{Fig1_supp}). We checked that at least one complete seasonal cycle was available for each station. Regional winds were extracted at the same locations and periods from the ERA5-Land dataset, which is a replay at a smaller spatial resolution of ERA5, the latest climate reanalysis from the ECMWF \citep{Hersbach2020, munoz2021}. This dataset provided hourly predictions of the 10-m wind velocity and direction at a spatial resolution of $0.1^\circ\times0.1^\circ$ ($\simeq 9~\textup{km}$ in Namibia).

To enable direct comparison, the local wind measurements were averaged into 1-hr bins centered on the temporal scale of the ERA5-Land estimates (Online Resource Fig.~\ref{Fig2_supp}). As the wind velocities of both datasets were provided at different heights, we converted them into shear velocities $u_{*}$ (Online Resource section 1), characteristic of the turbulent wind profile. Wind roses in Fig.~\ref{Fig2} show the resulting wind data.

Dune properties were computed using autocorrelation on the 30-m Digital Elevation Models (DEMs) of the shuttle radar topography mission \citep{Farr2007}. For the North and South Sand Sea stations, we obtain, respectively, orientations of $85^\circ$ and $125^\circ$ with respect to the North, wavelengths of $2.6~\textup{km}$ and $2.3~\textup{km}$ and amplitudes (or half-heights) of $45~\textup{m}$ and $20~\textup{m}$ (Online Resource Fig.~\ref{Fig4_supp} for more details). This agrees with direct measurements made on site.

\begin{figure}
\centering
\includegraphics[scale=1]{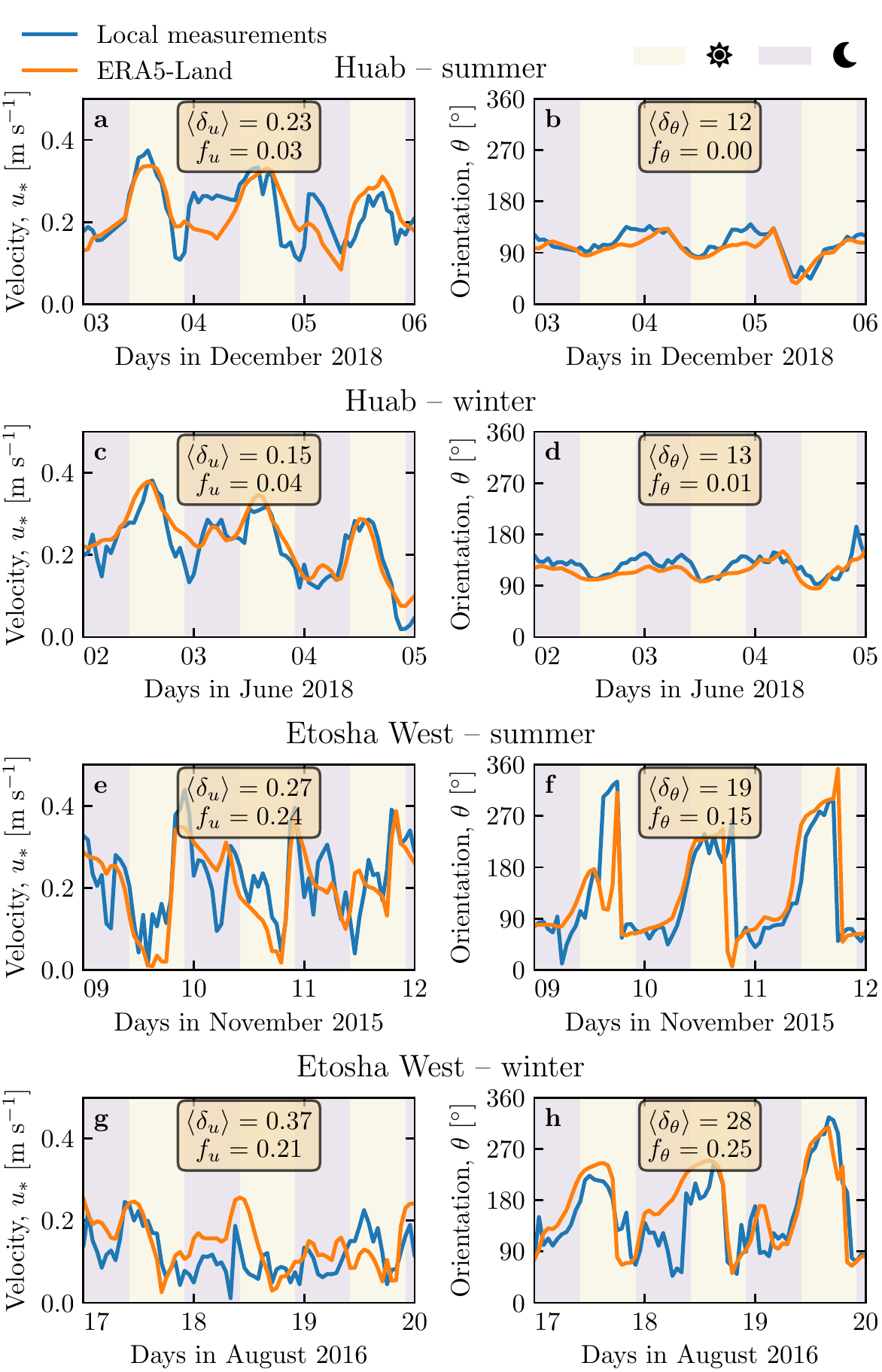}
\caption{Temporal comparison between the wind data coming from the ERA5-Land climate reanalysis (orange lines) and from the local measurements (blue lines). Coloured swathes indicate day (between 10.00 UTC and 22.00 UTC) and night (before 10.00 UTC or after 22.00 UTC). Numbers in legends indicate the average flow deflection $\delta_{\theta}$ and relative wind modulation $\delta_{u}$ over the displayed period (see Sect.~\ref{DataDistribution} for their definitions), as well as the percentage $f_\theta$ and $f_u$ of occurrence of extreme events ($\delta_{\theta} > 50^\circ$, $\vert\delta_{u}\vert > 0.6$). \textbf{a,b}: Huab station in summer. \textbf{b,c}: Huab station in winter. \textbf{d,e}: Etosha West station in summer. \textbf{f,g}: Etosha West station in winter. Time series of the two other stations are shown in Fig.~\ref{Fig5}.}
\label{Fig3}
\end{figure}

\begin{figure}
\centering
\includegraphics[scale=1]{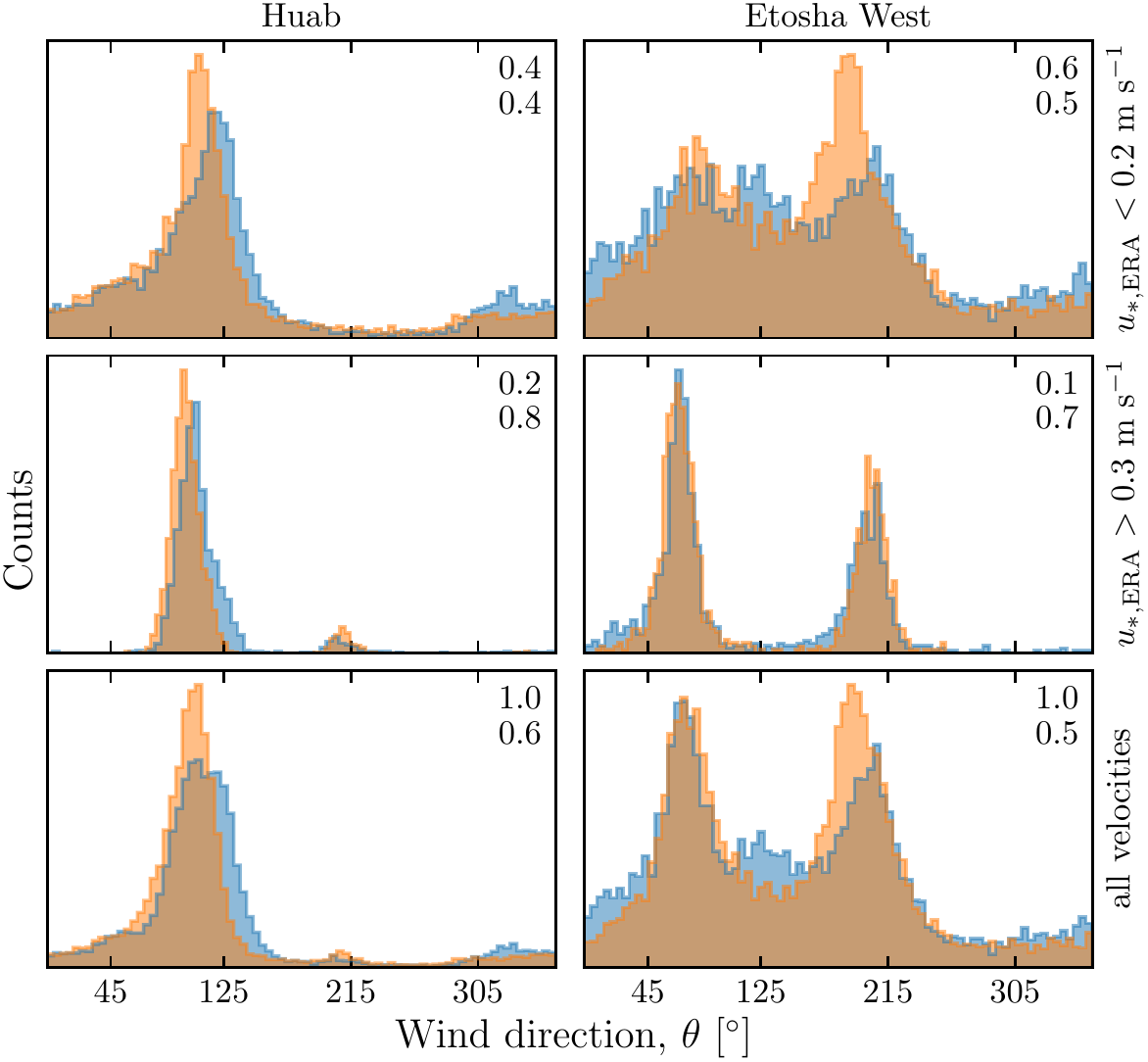}
\caption{Distributions of wind direction at Huab and Etosha West stations for the ERA5-Land climate reanalysis (orange) and the local measurements (blue). In each subplot, both distributions are plotted from the same time steps, selected for different ranges of the wind wind velocity (rows) in the ERA5-Land dataset. The numbers at the upper right corners give the percentage of time steps selected in each sub-range (top), as well as the percentage of them corresponding to the day -- defined between 10.00 UTC and 22.00 UTC (bottom).}
\label{Fig4}
\end{figure}

\subsection{Comparison of Local and Regional Winds}
\label{section_data_feedback}
The measured and predicted wind regimes are shown in Fig.~\ref{Fig2}. In the Namib, the regional wind patterns are essentially controlled by the sea breeze, resulting in strong northward components (sometimes slightly deviated by the large scale topography) present in all regional wind roses \citep{lancaster1985}. These daytime winds are dominant during the period October--March (Fig.~\ref{Fig3}f and Online Resource Fig.~\ref{Fig4}f). During April--September, an additional (and often nocturnal) easterly component can also be recorded, induced by the combination of katabatic winds forming in the mountains, and infrequent `berg' winds, which are responsible for the high wind velocities observed \citep{lancaster1984}. The frequency of these easterly components decreases from inland to the coast. As a result, bidirectional wind regimes within the Namib Sand Sea and at the west Etosha site (Fig.~\ref{Fig2}a,c and d) and a unidirectional wind regime on the coast at the outlet of the Huab River (Fig.~\ref{Fig2}b) are observed.

In the case of the Etosha West and Huab stations, the time series of wind speed and direction from the regional predictions quantitatively match those corresponding to the local measurements (Figs.~\ref{Fig3}, ~\ref{Fig4} and Online Resource Fig.~\ref{Fig5_supp}). For the North Sand Sea and South Sand Sea stations within the giant linear dune field, we observe that this agreement is also good, but limited to the October-March time period (Fig.~\ref{Fig4}a, b and e, f). However, the field-measured wind roses exhibit additional wind components aligned with the dune orientation, as evidenced on the satellite images (Fig.~\ref{Fig2}c and d).

\begin{figure}
\centering
\includegraphics[scale=1]{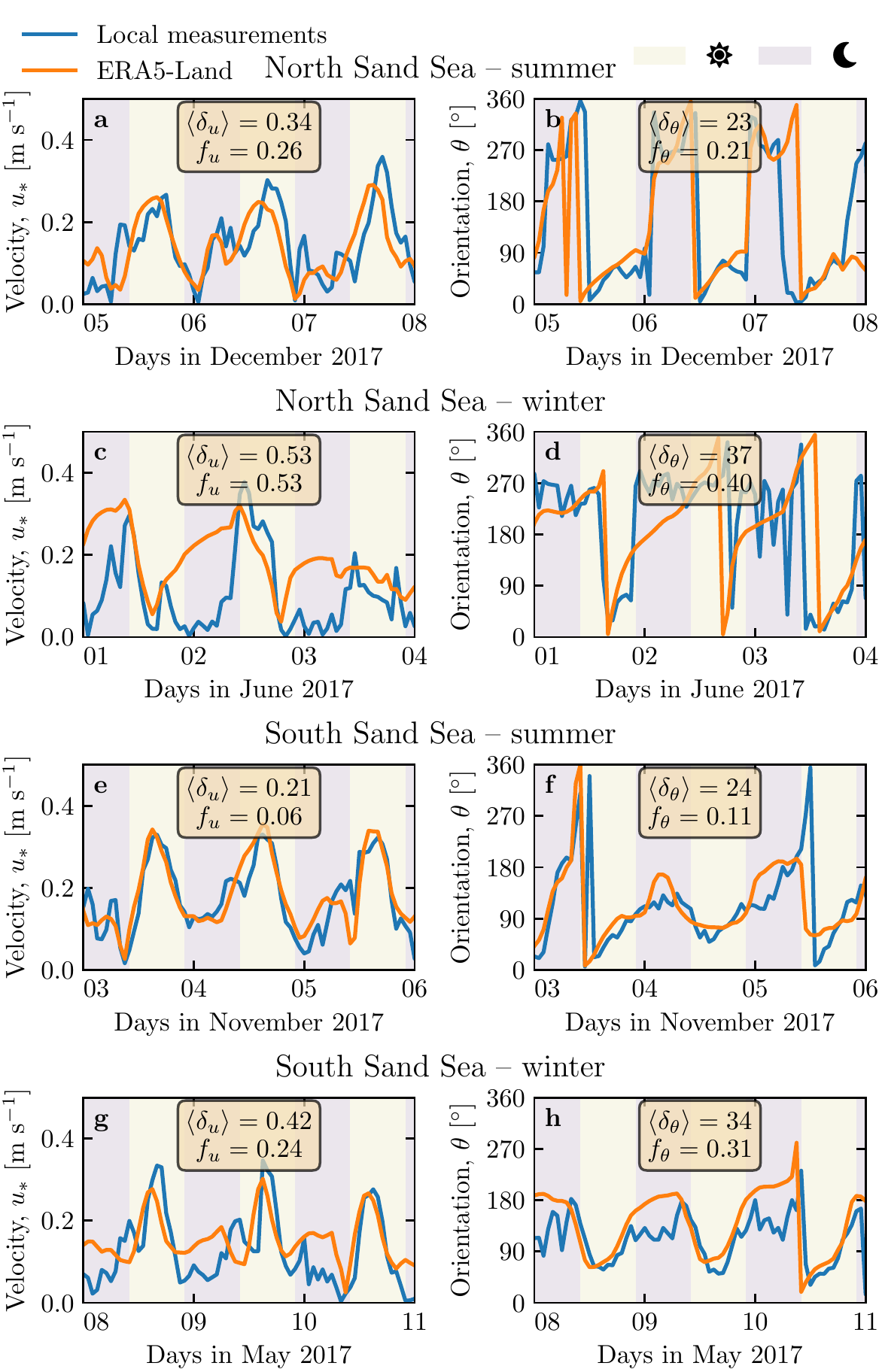}
\caption{Same as Fig.~\ref{Fig3} for North Sand Sea station in summer (\textbf{a,b}), North Sand Sea station in winter (\textbf{b,c}), South Sand Sea station in summer (\textbf{d,e}) and South Sand Sea station in winter (\textbf{f,g}).}
\label{Fig5}
\end{figure}

\begin{figure}
\centering
\includegraphics[scale=1]{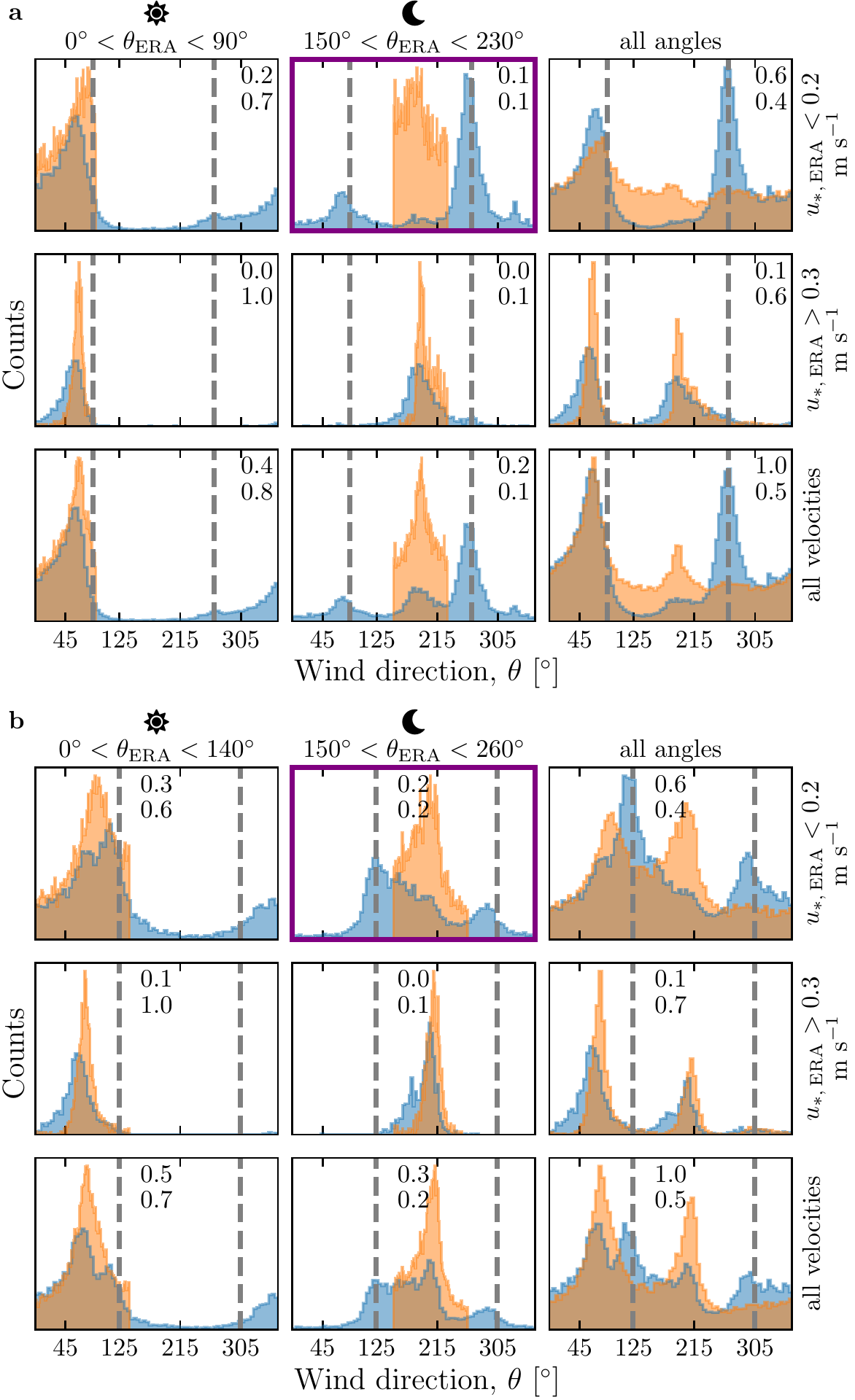}
\caption{Same as Fig.~\ref{Fig4} but for North Sand Sea (\textbf{a}) and South Sand Sea (\textbf{b}) stations. Here, subplots correspond to different ranges for the wind direction (columns) and wind velocity (rows) of the ERA5-Land dataset. The grey vertical dashed lines indicate the main dune orientation. In contrast with observations at the Huab and Etosha West stations (Fig.~\ref{Fig4}), histograms do not match well at low wind velocities, and the purple frame highlights the regime (low wind velocities, nocturnal easterly wind) in which the data from both datasets differ most.}
\label{Fig6}
\end{figure}

More precisely, during the April--September period, the local and regional winds in the interdune match during daytime only, i.e when the southerly/southwesterly sea breeze dominates (Figs.~\ref{Fig5}c,d,g,h and \ref{Fig6}). In the late afternoon and during the night, when the easterly `berg' and katabatic winds blow, measurements and predictions differ. In this case, the angular wind distribution of the local measurements exhibits two additional modes corresponding to reversing winds aligned with the dune orientation (purple frame in Fig.~\ref{Fig6}, Online Resource Fig.~\ref{Fig6_supp}). This deviation is also associated with a general attenuation of the wind strength (Online Resource Fig.~\ref{Fig7_supp}). Remarkably, all these figures show that these wind reorientation and attenuation processes occur only at low velocities of the regional wind, typically for $u_{*}^{\textup{ERA5-Land}} \lesssim 0.2~\textrm{m}~\textrm{s}^{-1}$. For shear velocities larger than $u_{*}^{\textup{ERA5-Land}} \simeq 0.3~\textrm{m}~\textrm{s}^{-1}$, the wind reorientation is not apparent. Finally, for intermediate shear velocities, both situations of wind flow reoriented along the dune crest and not reoriented can be successively observed (Online Resource Fig.~\ref{Fig6_supp}). Importantly, these values are not precise thresholds (and certainly not related to the threshold for sediment transport), but indicative of a crossover between regimes, whose physical interpretation is discussed in the next section.

\section{Influence of Wind Speed and Circadian Cycle on the Atmospheric Boundary Layer}
\label{sec:sec3}
The wind deflection induced by dunes has previously been related to the incident angle between wind direction and crest orientation, with a maximum deflection evident for incident angles between $30^{\circ}$ and $70^{\circ}$ \citep{Walker2009, Hesp2015}. In the data analysed here, the most deflected wind at both the North and South Sand Sea stations is seen to be where the incident angle is perpendicular to the giant dunes (Figs.~\ref{Fig2} and \ref{Fig6}). It therefore appears that in our case, the incident wind angle is not the dominant control on maximum wind deflection. Further, and as shown in Fig.~\ref{Fig6}, winds of high and low velocities show contrasting behaviour in characteristics of deflection. This suggests a change in hydrodynamical regime between the winds. In this section, we discuss the relevant parameters associated with the dynamical mechanisms that govern the interactions between the atmospheric boundary layer flow and giant dune topographies. This analysis allows us to provide a physics-based interpretation of our measured wind data.

\begin{figure}
\centering
\includegraphics[scale=1]{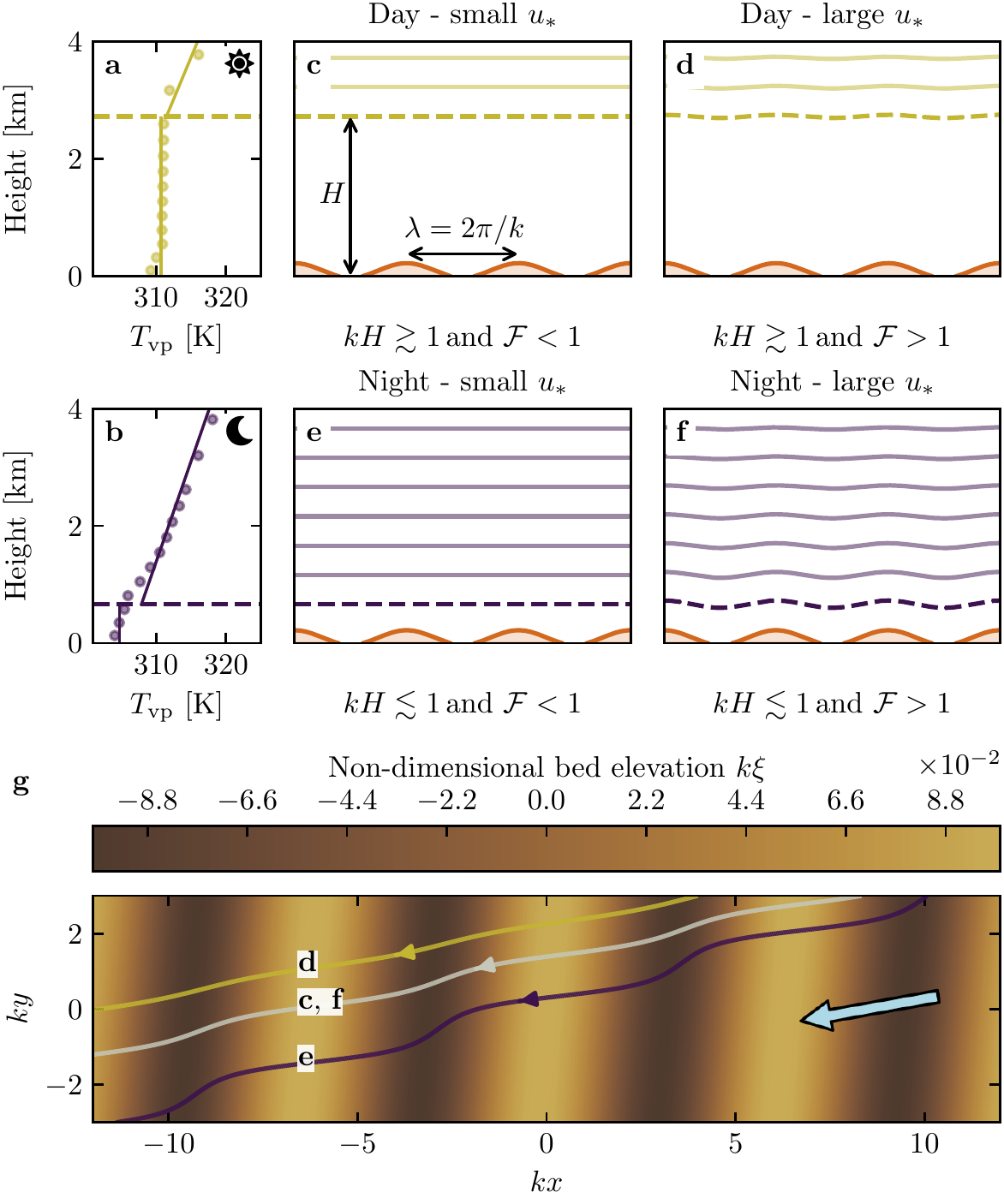}
\caption{\textbf{a,b}: Vertical profiles of the virtual potential temperature $T_{\rm vp}$ at two different time steps (day - 03/11/2015 - 12.00 UTC, night - 01/13/2013 - 09.00 UTC) at the North Sand Sea station. Dots: data from the ERA5 reanalysis. Dashed lines: boundary layer height given by the ERA5 reanalysis (Online Resource section 2). Plain lines: vertical (boundary layer) and linear (free atmosphere) fits to estimate the stratification properties. \textbf{c--f}: Sketches representing the interaction between the giant dunes and the atmospheric flow for different meteorological conditions. \textbf{g}: Streamlines over a sinusoidal topography $\xi(x,y)$ qualitatively representing the effect of low, medium and strong flow confinement, in relation to the above panels (see Appendix~\ref{turbulent_wind_model} for more details). The blue arrow indicates the undisturbed wind direction.}
\label{Fig7}
\end{figure}

\subsection{Flow Over a Modulated Bed}
\label{theoretical_framework}
Taking as a reference the turbulent flow over a flat bed, the general framework of our study is understanding and describing the flow response to a bed modulation (e.g. a giant dune). Without loss of generality, we can consider in this context an idealised bed elevation in the form of parallel sinusoidal ridges, with wavelength $\lambda$ (or wavenumber $k = 2\pi/\lambda$) and amplitude $\xi_0$, and where the reference flow direction makes a given incident angle with respect to the ridge crest \citep{Andreotti2012}. Part of this response, on which we focus here, is the flow deflection by the ridges. In a simplified way, it can be understood from the Bernoulli principle~\citep{Hesp2015}: as the flow approaches the ridge crest, the compression of the streamlines results in larger flow velocities, and thus lower pressures \citep{Jackson1975}. An incident flow oblique to the ridge is then deflected towards lower pressure zones, i.e towards the crest. Turbulent dissipation tends to increase this effect downstream, resulting in wind deflection along the crest in the lee side~\citep{Gadal2019}.

Flow confinement below a capping surface, which enhances streamline compression, has a strong effect on the hydrodynamic response and typically increases flow deflection. This is the case for bedforms forming in open channel flows such as rivers~\citep{Kennedy1963, Chang1970, Mizumura1995, Colombini2004, Fourriere2010, Andreotti2012, Unsworth2018}. This is also relevant for aeolian dunes as they evolve in the turbulent atmospheric boundary layer (ABL) capped by the stratified free atmosphere (FA) \citep{Andreotti2009}. Two main mechanisms, associated with dimensionless numbers must then be considered (Fig.~\ref{Fig7}). First, topographic obstacles typically disturb the flow over a characteristic height similar to their length. As flow confinement is characterised by a thickness $H$, the interaction between the dunes and the wind in the ABL is well captured by the parameter $k H$. The height $H$ is directly related to the sensitive heat flux from the Earth surface. It is typically on the order of a kilometre, but significantly varies with the circadian and seasonal cycles. Emerging and small dunes, with wavelengths in the range $20$ to $100$~m, are not affected by the flow confinement, corresponding to $k H \gg 1$. For giant dunes with kilometer-scale wavelengths, however, their interaction with the FA can be significant \citep{Andreotti2009}. This translates into a parameter $kH$ in the range $0.02$--$5$, depending on the moment of the day and the season. A second important mechanism is associated with the existence of a thin intermediate so-called capping layer between the ABL and the FA. It is characterised by a density jump $\Delta\rho$, which controls the `rigidity' of this interface, i.e. how much its deformation affects streamline compression. This is usually quantified using the Froude number \citep{Vosper2004, Stull2006, Sheridan2006, Hunt2006, Jiang2014}:
\begin{equation}
\mathcal{F} = \displaystyle\frac{U}{\sqrt{\displaystyle\frac{\Delta\rho}{\rho_{0}}gH}},
\label{FroudeNumber}
\end{equation}
where $U$ is the wind velocity at the top of the ABL and $\rho_{0}$ its average density. The intensity of the stratification, i.e. the amplitude of the gradient $\left | \partial_z \rho \right|$ in the FA, also impacts the ability to deform the capping layer under the presence of an underlying obstacle, and thus affects the influence of flow confinement. This can be quantified using the internal Froude number~\citep{Vosper2004, Stull2006, Sheridan2006, Hunt2006, Jiang2014} $\mathcal{F}_{\textup{I}} = k U/N$, where $N = \sqrt{-g \partial_z \rho / \rho_{0}}$ is the the Brunt-V\"ais\"al\"a frequency \citep{Stull1988}. Both Froude numbers have in practice the same qualitative effect on flow confinement (a smaller Froude corresponding to a stiffer interface), and we shall restrict the main discussion to $\mathcal{F}$ only.

With this theoretical framework in mind, and in the context of the measured wind data in the North and South Sand Sea stations, the smallest wind disturbances are expected to occur during the day, when the ABL depth is the largest and comparable to the dune wavelength ($k H \gtrsim 1$), which corresponds to a weak confinement situation (Fig.~\ref{Fig7}c and d). In contrast, large wind disturbances are expected to occur during the night, when the confinement is mainly induced by a shallow ABL (Fig.~\ref{Fig7}e). However, this strong confinement can be somewhat reduced in the case of strong winds, corresponding to large values of the Froude number and a less `rigid' interface (Fig.~\ref{Fig7}f). This is in qualitative agreement with the transition from deflected to non-deflected winds related to low and high velocities observed in our data (Sec.~\ref{section_data_feedback}).

\begin{figure}
\centering
\includegraphics[scale=1]{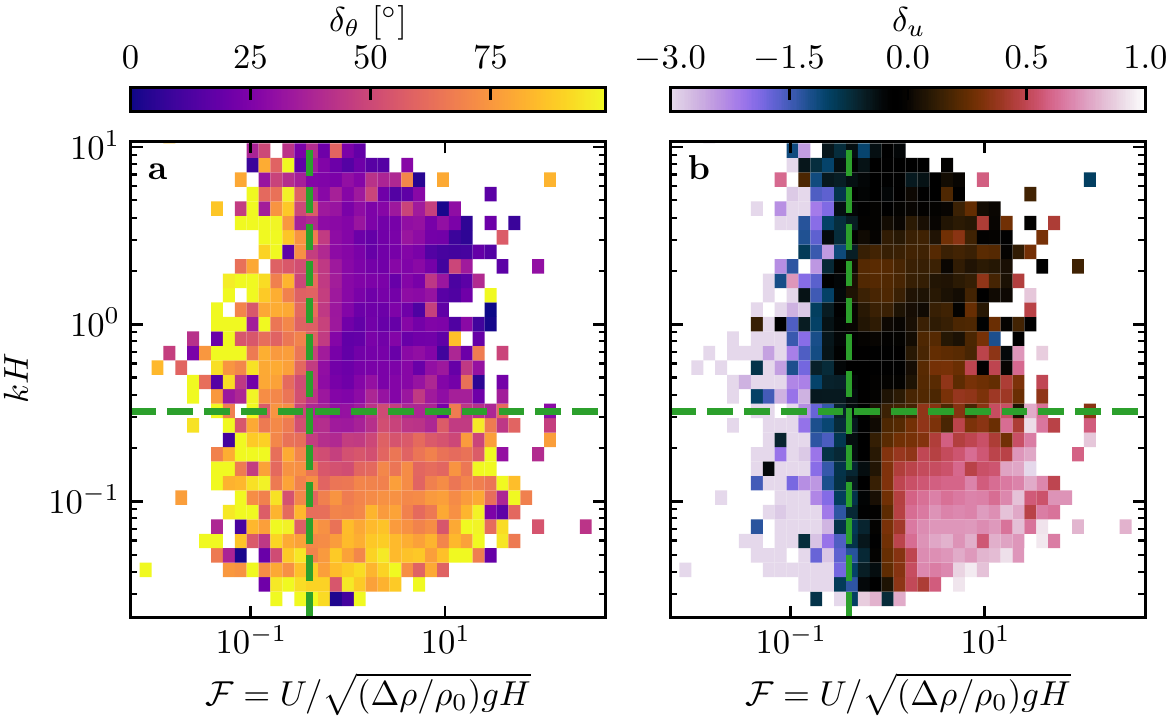}
\caption{Regime diagrams of the wind deviation $\delta_{\theta}$ (\textbf{a}) and relative attenuation/amplification $\delta_{u}$ (\textbf{b}) in the space $(\mathcal{F}, \, kH)$, containing the data from both the North Sand Sea and South Sand Sea stations. The green dashed lines empirically delimit the different regimes. The point density in each bin of the diagrams is shown in Online Resource Fig.~\ref{Fig10_supp} -- 95\% of the data occur in the range $-1 < \delta u < 1$. Similar regime diagrams in the spaces $(\mathcal{F}_{\textup{I}}, \, kH)$ and $(\mathcal{F}_{\textup{I}}, \, \mathcal{F})$ are shown in Online Resource Fig.~\ref{Fig11_supp}.}
\label{Fig8}
\end{figure}

\subsection{Data Distribution in the Flow Regimes}
\label{DataDistribution}
We can go one step further and analyse how our data quantitatively spread over the different regimes discussed above. For that purpose, one needs to compute $kH$ and $\mathcal{F}$ from the time series. $H$, $U$ and the other atmospheric parameters can be deduced from the various vertical profiles (temperature, humidity) available in the ERA5 climate reanalysis (Online Resource section 2). We quantify the flow deflection $\delta_{\theta}$ as the minimal angle between the wind orientations comparing the local measurements and the regional predictions. We also compute the relative velocity modulation as
\begin{equation}
\delta_{\textup{u}} = \frac{u_{*}^{\textup{ERA5-Land}} -  u_{*}^{\textup{Local mes.}}}{u_{*}^{\textup{ERA5-Land}}}.
\end{equation}
These two quantities are represented as maps in the plane $(\mathcal{F}, \, kH)$ (Fig.~\ref{Fig8}a and b), and one can clearly identify different regions in these plots. Small wind disturbances (small $\delta_{\theta}$ and $\delta_{\textup{u}}$) are located in the top-right part of the diagrams, corresponding to a regime with low-interaction as well as low-confinement ($k H$ and $\mathcal{F}$ large enough, Fig.~\ref{Fig7}d). Lower values of $k H$ (stronger interaction) or of Froude number (stronger confinement) both lead to an increase in wind disturbances, both in terms of orientation and velocity. Below a crossover value $k H \simeq 0.3$, wind disturbance is less sensitive to the $\mathcal{F}$-value. This is probably due to enhanced non-linear effects linked to flow modulation by the obstacle when confinement is strong (e.g. wakes and flow recirculations). The Froude number also controls a transition from damped to amplified wind velocities in the interdune, with a crossover around $\mathcal{F} \simeq 0.4$ (Fig.~\ref{Fig8}b). Such an amplification is rather unexpected. Checking the occurrence of the corresponding data, it appears that these amplifications are associated with the southerly sea breeze, and occur dominantly during the October-March period, when the other easterly wind is not present (Online Resource Fig.~\ref{Fig12_supp}a--b). Furthermore, they occur less frequently during the afternoon, and more frequently at the end of the day (Online Resource Fig.~\ref{Fig12_supp}c). This effect may be linked to a change in the flow behaviour in the lee side of the obstacles but further measurements are needed in order to assess the different possibilities \citep{baines1995, Vosper2004}.

As the hydrodynamic roughness $z_0$ determine the magnitude of wind shear velocities, Froude
number $\mathcal{F}$ and relative velocity modulation $\delta_{u}$, it is important to discuss
the sensitivity of the results to the $z_0$-values chosen for both the ERA5-Land and the field
data (see Online Resource Sect.~\ref{calib_z0}). Other quantities associated with wind direction are
independent of this choice. Considering the possible range of realistic roughness values, the
uncertainty on velocities estimated using the law of the wall is at most 30~\%. A similar
maximum uncertainty applies to the Froude number. This uncertainty also propagates to
$\delta_{u}$, for which Figure~\ref{Fig14_supp} shows that the choice of roughness has little influence on
its temporal variations,even if it can induce a global increase or decrease of its values. Hence, the choice
of the $z_0$-values will not qualitatively affect the overall aspect of the regime diagram
presented in Figure 8b. It may only change the value of $\delta_u$ for which the transition
between regimes is observed (dashed green lines in Figure~\ref{Fig8}b). Our conclusions are thus robust
with respect to the somewhat arbitrary choice of the hydrodynamic roughness values.

\begin{figure}
\centering
\includegraphics[scale=1]{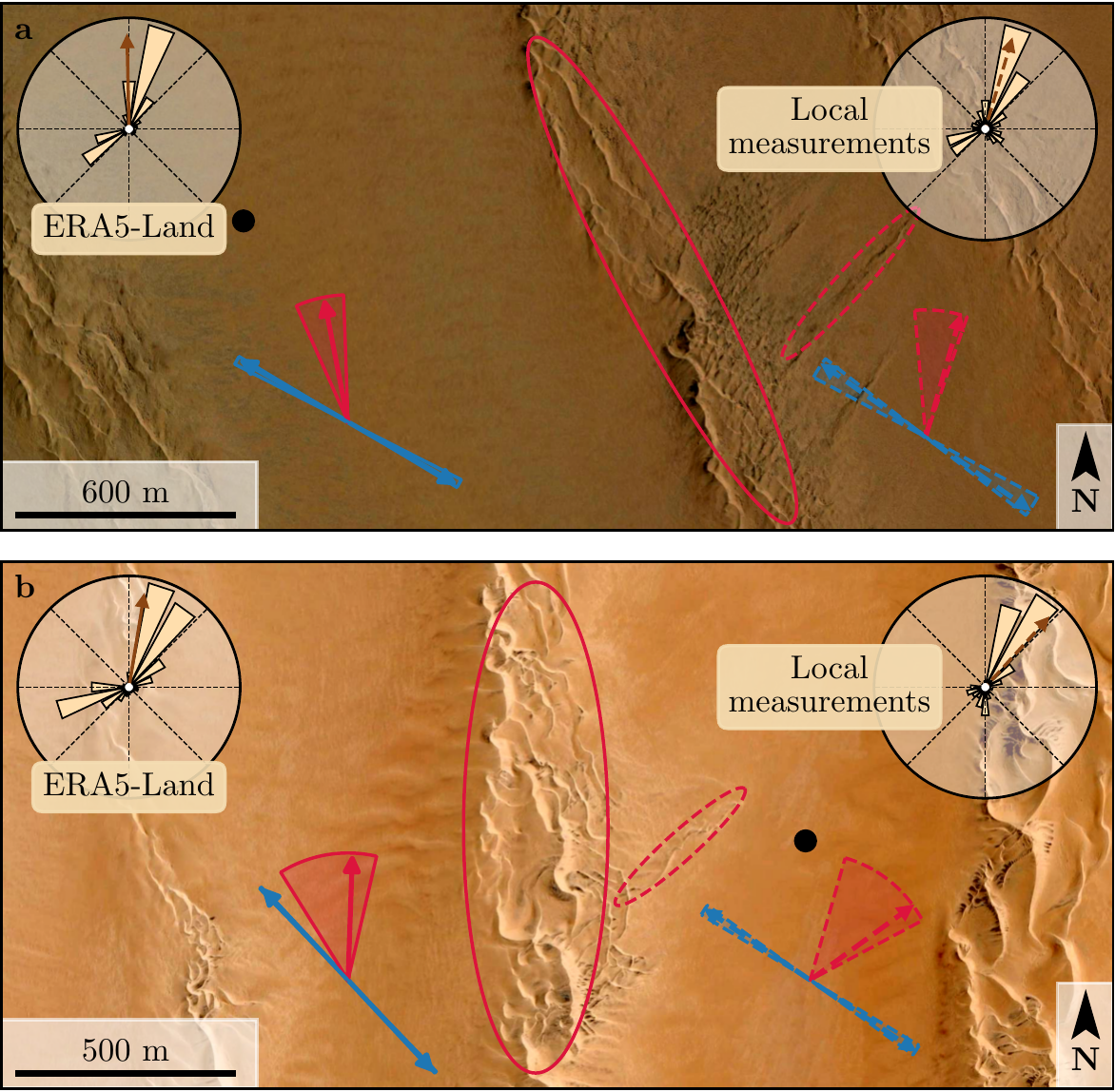}
\caption{Implications for smaller scale patterns in (\textbf{a}) the South Sand Sea and (\textbf{b}) North Sand Sea. The ellipses indicate the different types of elongating dunes, at large (plain line) and small (dashed line) scales. Dune orientation are predicted using the model of \citet{Courrech2014} from the sand flux angular distributions, shown here (roses) along with the resultant transport direction (brown arrow) for typical values (grain size $180~\mu$m, flux-up ratio of $1.6$). Code for corresponding arrows: use of ERA5-Land data (plain line), use of local measurements (dashed line), prediction for the bed instability growth mechanism (blue), prediction for the elongation growth mechanism (red). Wedges show the uncertainty on the orientation calculation when parameters are varied. The black dots indicate the location of the meteorological stations in the interdune. See Appendix~\ref{app:app2} for additional details.}
\label{Fig9}
\end{figure}

\section{Discussion and Conclusion}
\label{sec:sec4}

The feedback of the giant dunes on the wind flow has important implications for smaller scale bedforms. As illustrated in Fig.~\ref{Fig9}, small linear dunes ($\sim50$~m wide) are often present within the 1--2~km interdune spaces between giant linear dunes in the Namib Sand Sea \citep{Livingstone2010}. These smaller dunes do not exhibit the same orientation as the large ones, and are sometimes named `crossing dunes'~\citep{Chand22}. Whilst differences between large and small scale dune patterns are observed ubiquitously, they are usually attributed to the presence of two different dune growth mechanisms, leading to two different dune patterns (orientations and/or morphologies) for the same wind regime~\citep{Courrech2014, Runyon2017, lu2017, Song2019, Gadal2020, Hu2021}. Here, however, our arguments enable the development of differing orientations for the small and giant linear dunes governed by the same dune growth mechanism (elongating mode). Figure~\ref{Fig9} shows how the orientations for the small and giant dunes can be derived from the locally measured and regionally predicted winds respectively (red arrows in Fig.~\ref{Fig9}). These predictions require the threshold of aeolian sand transport to be specified. Importantly, its value (a shear velocity estimated at $u_{\textup{th}} \simeq 0.15~\textup{m}~\textup{s}^{-1}$ -- see Appendix~\ref{app:app2}) can be reached in periods during which deflected winds are observed (recall that the stronger winds, responsible for most of the sediment transport and associated dune morphodynamics, are not deflected -- see Fig.~\ref{Fig6}). The feedback of the giant dunes on the wind described in this study, through wind deflection and attenuation, thus provides a potential explanation for the existence of these small linear dunes elongating across the interdune, a dynamic which has remained unresolved to date. These crossing dunes could provide additional constraints for the inference of local winds from bedforms, similarly to that currently performed on Mars using ripple orientations \citep{Liu2015, Hood2021}. Further work is needed to investigate these processes in more detail, including measurements of sediment transport and flow on the top of dunes.

This study presents field and reanalysis-based evidence that wind flow patterns around giant dunes are influenced by the atmospheric boundary layer, particularly during nocturnal conditions. However, we do not address here the question of the limitation of the giant dune pattern coarsening, and leave open the debate as to whether their size is controlled by the depth of this layer \citep{Andreotti2009}, in contrast to sediment supply limited and ever-slower growth with size \citep{Werner1999, Gunn2022}. More field evidence is definitively needed from additional dune fields, but this mechanism would allow for the inference of the ABL depth from giant bedform wavelengths where measurements are not feasible or available, such as Titan \citep{Lorenz2010}.

To conclude on conditions under which the ERA5-Land reanalysis data can reliably be used to study dune morphodynamics, we summarise the comparison of local (direct measurements) and regional (climate reanalysis) wind data as follows. In flat areas, the agreement between the two confirms the ability of the ERA5-Land climate reanalysis to predict the wind regime down to scales $\sim10$~km, i.e the model grid. When smaller scale topographies are present (giant dunes in our case), locally measured winds can significantly differ from the regionally predicted ones. This is the case when the disturbances induced by the dunes interact with the lower part of the ABL vertical structure, which presents circadian variations. During the day, when the capping layer is typically high, this interaction is small, and the ERA5-Land predictions are also quantitatively consistent with the local data. During the night, however, the presence of a shallow atmospheric boundary layer induces a strong confinement of the flow, and is associated with large wind deflection by the dunes. Importantly, we find that this effect can be counterbalanced for large wind velocities, which are capable of deforming the capping layer, thus decreasing the influence of the confinement.

The theoretical computation of the wind disturbances induced by sinusoidal ridges under flow confinement has been performed in the linear limit \citep{Andreotti2009, Andreotti2012}, i.e. when the aspect ratio of these ridges is small ($k\xi_0 \ll 1$). These models are able to qualitatively reproduce the observed wind deflection (Appendix~\ref{turbulent_wind_model}, Online Resource Figs.~\ref{Fig11_supp} and \ref{Fig13_supp}), and thus provide the physical support for the interpretation we propose here based on hydrodynamic regimes. However, these models cannot quantitatively predict the magnitude of our observations, probably due to the presence of expected non-linearities in high confinement situations linked to strong flow modulations. Besides, these linear calculations only predict wind attenuation in the interdune, in contrast with the observed enhanced velocities associated with particular evening winds from the south during the period October-March (Online Resource Fig.~\ref{Fig12_supp}). Some other models predict different spatial flow structures in response to a modulated topography, such as lee waves and rotors \citep{baines1995, Vosper2004}. However, our measurements are located at a single point in the interdune, and we are thus unable to explore these types of responses. Data at different places along and across the ridges are needed to investigate and possibly map such flow structures, and for further comparisons with the models.

\begin{acknowledgements}
We would like to acknowledge the contributors of the following open-source python librairies, Matplotlib \citep{Hunter2007}, Numpy \citep{Harris2020} and Scipy \citep{Virtanen2020}, which provide an incredibly efficient ecosystem allowing scientific research in Python. We also thank B. Andreotti and A. Gunn for useful discussions.

\end{acknowledgements}

{\footnotesize\noindent\textbf{Data Availability}~All data used in this study can be found in \citet{DataPaper}. Note that it contains modified Copernicus Climate Change Service Information (2021). Neither the European Commission nor ECMWF is responsible for any use that may be made of the Copernicus Information or Data it contains. Documented codes used in this study to analyse this data are available at https://github.com/Cgadal/GiantDunes (will be made public upon acceptance of this manuscript for publication).}

\vspace{.2cm }

{\footnotesize\noindent\textbf{Fundings}~Multiple grants have supported the collection of wind data through visits to the four sites between 2013 and 2020 (John Fell Oxford University Press (OUP) Research Fund (121/474); National Geographic (CP-029R-17); Natural Environment Research Council UK (NE/R010196/1 and NE/H021841/1 NSFGEO-NERC); Southampton Marine and Maritime Institute SMMI EPSRC-GCRF UK), along with research permits (1978/2014, 2140/2016, 2304/2017, 2308/2017, RPIV00022018, RPIV0052018, RPIV00230218).  The authors are very grateful for support from Etosha National Park (especially Shyane K\"otting, Boas Erckie, Pierre du Preez, Claudine Cloete, Immanuel Kapofi, Wilferd Versfeld, and Werner Kilian), Gobabeb Namib Research Institute (Gillian Maggs-K\"olling and Eugene Marais), The Skeleton Coast National Park (Joshua Kazeurua).  Various researchers and desert enthusiasts have assisted with instruments and the logistics of expeditions, especially Mary Seely for expert guidance at the North Sand Sea site.
Finally, we acknowledge financial support from the Laboratoire d'Excellence UnivEarthS Grant ANR-10-LABX-0023, the Initiative d'Excellence Universit\'e de Paris Grant ANR-18-IDEX-0001, the French National Research Agency Grants ANR-17-CE01-0014/SONO and the National Science Center of Poland Grant 2016/23/B/ST10/01700.}

\vspace{.2cm }
{\footnotesize\noindent\textbf{Conflict of interest}~The authors declare that they have no conflict of interest.}

\section*{Appendix 1: Linear Theory of Wind Response to Topographic Perturbation}
\label{turbulent_wind_model}

Following the work of \citet{Fourriere2010}, \citet{Andreotti2012} and \citet{Andreotti2009}, we briefly describe in this appendix the framework for the linear response of a turbulent flow to a topographic perturbation of small aspect ratio. As a general bed elevation can be decomposed into Fourier modes, we focus here on a sinusoidal topography:
\begin{equation}
\xi = \xi_{0}\cos\left[k\left(\cos(\alpha)y - \sin(\alpha)x\right)\right],
\end{equation}
which is also a good approximation for the giant dunes observed in the North Sand Sea and South Sand Sea Station (Fig.~\ref{Fig2} and Online Resource Fig.~\ref{Fig4_supp}). Here, $x$ and $y$ are the streamwise and spanwise coordinates, $k=2\pi/\lambda$ the wavenumber of the sinusoidal perturbation, $\alpha$ its crest orientation with respect to the $x$-direction (anticlockwise) and $\xi_{0}$ its amplitude. The two components of the basal shear stress $\boldsymbol{\tau} = \rho_{0} u_{*}\boldsymbol{u}_{*}$, constant in the flat bottom reference case, can then be generically written as:
\begin{align}
\tau_{x} & = \tau_{0}\left(1 + k\xi_{0}\sqrt{\mathcal{A}_{x}^{2} + \mathcal{B}_{x}^{2}}\cos\left[k\left(\cos(\alpha)y - \sin(\alpha)x\right) + \phi_{x}\right]\right), \\
\tau_{y} & = \tau_{0} \, k\xi_{0}\sqrt{\mathcal{A}_{y}^{2} + \mathcal{B}_{y}^{2}}\cos\left[k\left(\cos(\alpha)y - \sin(\alpha)x\right) + \phi_{y}\right],
\end{align}
where $\tau_{0}$ is the reference basal shear stress on a flat bed. We have defined the phase $\phi_{x, y} = \tan^{-1}\left(\mathcal{B}_{x, y}/\mathcal{A}_{x, y}\right)$ from the in-phase and in-quadrature hydrodynamical coefficients $\mathcal{A}_{x, y}$ and $\mathcal{B}_{x, y}$. They are functions of $k$ and of the flow conditions, i.e the bottom roughness, the vertical flow structure and the incident flow direction, and the theoretical framework developed in the above cited papers proposes methods to compute them in the linear regime.

Following \citet{Andreotti2012}, the effect of the incident wind direction can be approximated by the following expressions:
\begin{align}
  \mathcal{A}_{x} & = \mathcal{A}_{0}\sin^{2}\alpha, \\
  \mathcal{B}_{x} & = \mathcal{B}_{0}\sin^{2}\alpha, \\
  \mathcal{A}_{y} & = -\displaystyle\frac{1}{2}\mathcal{A}_{0}\cos\alpha \sin\alpha, \\
  \mathcal{B}_{y} & = -\displaystyle\frac{1}{2}\mathcal{B}_{0}\cos\alpha \sin\alpha,
\end{align}
where $\mathcal{A}_{0}$ and $\mathcal{B}_{0}$ are now two coefficients independent of the dune orientation $\alpha$, corresponding to the transverse case ($\alpha=90$). In the case of a fully turbulent boundary layer capped by a stratified atmosphere, these coefficients depend on $k H$, $k z_{0}$, $\mathcal{F}$ and $\mathcal{F}_{\textup{I}}$ \citep{Andreotti2009}. For their computation, we assume here a constant hydrodynamic roughness $z_{0} \simeq 1$~mm (Online Resource section 1). For the considered giant dunes, this leads to $k z_{0} \simeq 2 \cdot 10^{-6}$, as their wavelength is $\lambda \simeq 2.4$~km (or $k \simeq 2 \cdot 10^{-3}$~m$^{-1}$). Values of $z_{0}$ extracted from field data indeed typically fall between 0.1 mm and 10 mm \citep{Sherman2008, Field2018}. Importantly, $\mathcal{A}_{0}$ and $\mathcal{B}_{0}$ do not vary much in the corresponding range of $k z_{0}$ \citep{Fourriere2010}, and the results presented here are robust with respect to this choice.

With capping layer height and Froude numbers computed from the ERA5-Land time series, the corresponding $\mathcal{A}_{0}$ and $\mathcal{B}_{0}$ can be deduced, as displayed in Online Resource Fig.~\ref{Fig13_supp}. Interestingly, it shows similar regimes as in the diagrams of Fig.~\ref{Fig8} and Online Resource Fig.~\ref{Fig11_supp}a,b, supporting the underlying physics. However, the agreement is qualitative only. Further, the linearity assumption of the theoretical framework requires $\left(\vert \tau \vert - \tau_{0}\right)/\tau_{0} \ll 1$, which translates into $k\xi\sqrt{\mathcal{A}_{0}^{2} + \mathcal{B}_{0}^{2}} \ll 1$. In our case, the giant dune morphology gives $k\xi_0 \simeq 0.1$, which means that one quits the regime of validity of the linear theory when the coefficient modulus $\sqrt{\mathcal{A}_{0}^{2} + \mathcal{B}_{0}^{2}}$ becomes larger than a few units. In accordance with the theoretical expectations, these coefficients present values on the order of unity ($\mathcal{A}_{0} \simeq 3$ and $\mathcal{B}_{0} \simeq 1$) in unconfined situations \citep{Claudin2013, Lu2021}. In contrast and as illustrated in Online Resource Fig.~\ref{Fig13_supp}a,b, larger values are predicted in case of strong confinement, which does not allow us to proceed to further quantitative comparison with the data.

Finally, the linear model is also able to reproduce the enhancement of the flow deflection over the sinusoidal ridges when $\sqrt{\mathcal{A}_{0}^{2} + \mathcal{B}_{0}^{2}}$ is increased (Online Resource Fig.~\ref{Fig13_supp}). Here, using $k\xi_0 \simeq 0.1$ to be representative of the amplitude of the giant dunes at the North Sand Sea station, the coefficient modulus is bounded to $10$.

\section*{Appendix 2: Sediment Transport and Dune Morphodynamics}
\label{app:app2}

We summarise in this appendix the sediment transport and dune morphodynamics theoretical framework leading to the prediction of sand fluxes and dune orientations from wind data.

\paragraph{Sediment Transport} ---
The prediction of sand fluxes from wind data has been a long standing issue in aeolian geomorphological studies~\citep{Fryberger79, Pearce2005, Sherman2012, Shen2019}. Based on laboratory studies in wind tunnels \citep{Rasmussen96, Iversen99, Creyssels2009, Ho2011}, as well as physical considerations \citep{Ungar1987, Andreotti2004bis, Duran2011, Pahtz2020}, it has been shown that the steady saturated saltation flux over a flat sand bed depends linearly on the shear stress:
\begin{equation}
\label{transport_law_quadratic}
\frac{q_{\textup{sat}}}{Q} = \Omega\sqrt{\Theta_{\textup{th}}}\left(\Theta - \Theta_{\textup{th}}\right),
\end{equation}
where $\Omega$ is a proportionality constant, $Q = d\sqrt{(\rho_{\textup{s}} - \rho_{0})gd/\rho_{0}}$ is a characteristic flux, $\Theta = \rho_{0} u_{*}^{2}/(\rho_{\textup{s}} - \rho_{0})gd$ the Shields number, and $\Theta_{\textup{th}}$ its threshold value below which saltation vanishes. $\rho_{\textup{s}} = 2.6~\textup{g}~\textup{cm}^{-3}$ and $d=180~\mu\textup{m}$ are the grain density and diameter, and $g$ is the gravitational acceleration. The shear velocity, and consequently the Shields number as well as the sediment flux, are time dependent.

Recently, \citet{Pahtz2020} suggested an additional quadratic term in Shields to account for grain-grain interactions within the transport layer at strong wind velocities:
\begin{equation}
\label{transport_law_quartic}
\frac{q_{\textup{sat, t}}}{Q} = \frac{2\sqrt{\Theta_{\textup{th}}}}{\kappa\mu}\left(\Theta - \Theta_{\textup{th}}\right)\left(1 +\frac{C_{\textup{M}}}{\mu}\left[\Theta - \Theta_{\textup{th}}\right]\right),
\end{equation}
where $\kappa = 0.4$ is the von K\'arm\'an constant, $C_{\rm M} \simeq 1.7$ a constant and $\mu \simeq 0.6$ is a friction coefficient, taken to be the avalanche slope of the granular material. The fit of this law to the experimental data of \citet{Creyssels2009} and \citet{Ho2011} gives $\Theta_{\textup{th}} = 0.0035$. The fit of Eq.~\ref{transport_law_quadratic} on these same data similarly gives $\Omega \simeq 8$ and $\Theta_{\textup{th}} = 0.005$. The sand flux angular distributions and the dune orientations in Fig.~\ref{Fig9} are calculated using this law \eqref{transport_law_quartic}. We have checked that using the ordinary linear relationship \eqref{transport_law_quadratic} instead does not change the predicted dune orientations by more than a few degrees.

\paragraph{Dune Orientations} ---
Dune orientations are predicted with the dimensional model of \citet{Courrech2014}, from the sand flux time series computed with the above transport law. Two orientations are possible depending on the mechanism dominating the dune growth: elongation or bed instability. The orientation $\alpha$ corresponding to the bed instability is then the one that maximises the following growth rate \citep{Rubin1987}:
\begin{equation}
\sigma \propto \frac{1}{H_{d} W_{d} T}\int_0^T  q_{\textup{crest}}\vert \sin\left(\theta - \alpha\right) \vert \, \textup{d}t,
\end{equation}
where $\theta$ is the wind orientation measured with respect to the same reference as $\alpha$, and $H_{d}$ and $W_{d}$ are dimensional constants respectively representing the dune height and width. The integral runs over a time $T$, which must be representative of the characteristic period of the wind regime. The flux at the crest is expressed as:
\begin{equation}
q_{\textup{crest}} = q_{\textup{sat}}\left[1 + \gamma\vert\sin\left(\theta - \alpha\right)\vert\right],
\end{equation}
where the flux-up ratio $\gamma$ has been calibrated to $1.6$ using field studies, underwater laboratory experiments and numerical simulations. Predictions of the linear analysis of \citet{Gadal2019} and \citet{Delorme2020} give similar results.

Similarly, the dune orientation corresponding to the elongation mechanism is the one that verifies:
\begin{equation}
\tan(\alpha) = \frac{\langle q_{\textup{crest}}(\alpha) \boldsymbol{e}_{\theta}\rangle \cdot \boldsymbol{e}_{WE} }{ \langle q_{\textup{crest}}(\alpha) \boldsymbol{e}_{\theta} \rangle \cdot \boldsymbol{e}_{SN}},
\end{equation}
where $\langle.\rangle$ denotes a vectorial time average. The unitary vectors $\boldsymbol{e}_{WE}$, $\boldsymbol{e}_{SN}$ and $\boldsymbol{e}_{\theta}$ are in the West-East, South-North and wind directions, respectively.

The resulting computed dune orientations, blue and red arrows in Fig.~\ref{Fig9}, then depend on a certain number of parameters (grain properties, flux-up ratio, etc.), for which we take typical values for aeolian sandy deserts. Due to the lack of measurements in the studied places, some uncertainties can be expected. We therefore run a sensitivity test by calculating the dune orientations for grain diameters ranging from $100~\mu\textup{m}$ to $400~\mu\textup{m}$ and for a speed-up ratio between $0.1$ and $10$ (wedges in Fig.~\ref{Fig9}).

\clearpage

\bibliographystyle{spbasic_updated}
\bibliography{main}


\newpage

\renewcommand{\thefigure}{S\arabic{figure}}
\setcounter{figure}{0}

\begin{center}
\textbf{\large
Local wind regime induced by giant linear dunes \\
--- Supplementary Material ---
}
\end{center}

\noindent
\textbf{C. Gadal$^{*}$ $\cdot$ P. Delorme $\cdot$ C. Narteau $\cdot$ G.F.S. Wiggs $\cdot$ M. Baddock $\cdot$ J.M. Nield $\cdot$ P. Claudin}
\\ \\
$^{*}$ Institut de M\'ecanique des Fluides de Toulouse, Universit\'e de Toulouse Paul Sabatier, CNRS, Toulouse INP-ENSEEIHT, Toulouse, France.\\
\texttt{cyril.gadal@imft.fr}

\section*{1. Shear velocity and calibration of the hydrodynamical roughness}
\label{calib_z0}

As the regionally predicted and locally measured velocities are available at different heights, we can not compare them directly. We therefore convert all velocities into shear velocities $u_{*}$, characteristic the turbulent logarithmic velocity profile \citep{Spalding1961, Stull1988}:
\begin{equation}
\frac{u(z)}{u_{*}} = \frac{1}{\kappa}\ln\left(\frac{z}{z_{0}}\right),
\end{equation}
where $z$ is the vertical coordinate, $\kappa = 0.4$ the von K\'arm\'an constant and $z_{0}$ the hydrodynamic roughness. Note that, strickly speaking, this logarithmic profile is valid for a neutrally stratified ABL only. Vertical density gradients occuring in other conditions may thus induce large discrepancies~\citep{Monin1954, Garratt1994, Dyer1974}. However, as our wind measurements are in the flow region close enough to the surface, where these effects are negligible, this logarithmic wind profile remains a farily good approximation in all conditions~\citep{gunn2021}.
Several measurements of hydrodynamic roughnesses are available \citep{Raupach1992,Bauer1992,Brown2008,Nield2014}. In the absence of sediment transport, it is governed by the geometric features of the bed \citep{Flack2010,Pelletier2016}. When aeolian saltation occurs, it is rather controlled by the altitude of Bagnold's focal point \citep{Duran2011,Valance2015}, which depends on the wind velocity and grain properties \citep{Sherman2008, Zhang2016, Field2018}. Whether associated with geometric features or with sediment transport, its typical order of magnitude is the millimetre scale on sandy surfaces.

We do not have precise velocity vertical profiles to be able to deduce an accurate value of $z_0$ in the various environments of the meteorological stations (vegetated, arid, sandy). Our approach is to rather select the hydrodynamic roughness which allows for the best possible matching between the regionally predicted and locally measured winds, i.e. minimising the relative difference $\delta$ between the wind vectors of both datasets:
\begin{equation}
\label{metric_roughness}
\delta = \frac{\sqrt{\langle\| \boldsymbol{u}_{*, \textrm{era}} - \boldsymbol{u}_{*, \textrm{station}} \|^{2}\rangle}}{\sqrt{ \langle \| \boldsymbol{u}_{*, \textrm{era}} \| \rangle \langle \| \boldsymbol{u}_{*, \textrm{station}} \| \rangle}} ,
\end{equation}
where $\langle.\rangle$ denotes time average. This parameter is computed for values of $z_0$ in ERA5-Land analysis ranging from $10^{-5}$~m to $10^{-2}$~m for the four different stations. Note that for the North Sand Sea and South Sand Sea stations, where the giant dunes feedback presumably affect the wind, we take into account the non-deflected winds only in the calculation of $\delta$ (with a $15^\circ$ tolerance).

As shown in Online Resource Fig.~\ref{Fig3_supp}, the minimum values of $\delta$ in the space ($z_{0}^{\textup{ERA5Land}}$, $z_{0}^{\textup{local}}$) form a line. We thus set the roughness in the ERA5-Land analysis to the typical value $z_0=10^{-3}$~m, and deduce the corresponding ones for the local stations. It leads to $2.7$, $0.8$, $0.1$ and $0.5$~mm for the Etosha West, North Sand Sea, Huab and South Sand Sea stations, respectively. Importantly, this approach somewhat impacts the calculation of the shear velocities, but not that of the wind directions. As such, most of our conclusions are independent of such a choice. However, it may affect the magnitude of the wind velocity attenuation/amplification in flow confinement situations.

\section*{2. Computation of the ABL characteristics}

The estimation of the non-dimensional numbers associated with the ABL requires the computation of representative meteorological quantities. In arid areas, the vertical structure of the atmosphere can be approximated by a well mixed convective boundary layer of height $H$, topped by the stratified free atmosphere \citep{Stull1988, Shao2008}. In this context, one usually introduces the virtual potential temperature $T_{\textup{vp}}$, which is a constant $T_{0}$ inside the boundary layer, and increases linearly in the FA (Online Resource Fig.~\ref{Fig8_supp}a):
\begin{equation}
T_{\textup{vp}}(z) =
\begin{cases}
T_{0} &\text{for $z \leq H$},\\
T_{0}\left(1 + \displaystyle\frac{\Delta T_{\textup{vp}}}{T_{0}} + \frac{N^{2}}{g}(z - H)\right) &\text{for $ z \geq H$},
\end{cases}
\label{TvpVerticalProfile}
\end{equation}
where $\Delta T_{\textup{vp}}$ is the temperature discontinuity at the capping layer and $N = \sqrt{g \partial_z T_{\textup{vp}} / T_{0}}$ is the Brunt-V\"ais\"al\"a frequency, characteristic of the stratification. Note that, under the usual Boussinesq approximation, temperature and air density variations are simply related by $\delta T_{\textup{vp}}/T_{0} \simeq - \delta\rho/\rho_{0}$ (see Online Resource of \citet{Andreotti2009}), so that $N$ can equivalently be defined from the density gradient as next to \eqref{FroudeNumber}.

The ERA5 dataset provides vertical profiles of the geopotential $\phi$, the actual temperature $T$ and the specific humidity $\eta$ at given pressure levels $P$. The vertical coordinate is then calculated as:
\begin{equation}
z = \frac{\phi R_{\textup{t}}}{g R_{\textup{t}} - \phi},
\end{equation}
where $R_{\textup{t}} = 6371229$~m is the reference Earth radius and $g = 9.81$~m s$^{-2}$ is the gravitational acceleration. One also computes the virtual potential temperature as:
\begin{equation}
T_{\textup{vp}} = T\left[1 + \left(\frac{M_d}{M_w} - 1\right) \eta \right] \left(\frac{P_{0}}{P}\right)^{R/C_p},
\end{equation}
where $P_{0} = 10^{5}$~Pa is the standard pressure, $R = 8.31$~J/K is the ideal gas constant, $C_p \simeq 29.1$~J/K is the air molar heat capacity, and $M_w = 0.018$~kg/Mol and $M_d = 0.029$~kg/Mol are the molecular masses of water and dry air respectively. The specific humidity is related to the vapour pressure $p_w$ as
\begin{equation}
\eta = \frac{\frac{M_w}{M_d} p_w}{p - \left( 1 - \frac{M_w}{M_d} \right) p_w} \, .
\end{equation}

The ERA5 dataset also provides an estimate of the ABL depth $H$, based on the behaviour of the Richardson vertical profile. This dimensionless number is defined as the ratio of buoyancy and flow shear terms, and can be expressed as $\textup{Ri} = N^2/(\partial_z u)^2$. It vanishes in the lower well-mixed layer where $T_{\textup{vp}}$ is constant, and increases in the stratified FA. Following the method and calibration of \citet{Vogelezang1996,seidel2012}, the value $\textup{Ri}(z) \simeq 0.25$ has been shown to be a good empirical criterion to give $z \simeq H$ within a precision varying from $50$\% for the shallower ABL (e.g. at night) to $20$\% for situations of stronger convection.

Examples of vertical virtual potential temperature profiles deduced from ERA5 are shown in Online Resource Fig.~\ref{Fig8_supp}a. For each of them, an average temperature is computed below the ABL depth ($z<H$), and a linear function is fitted above, allowing us to extract the temperature jump $\Delta T_{\textup{vp}}$. Importantly, some profiles display a vertical structure that cannot be approximated by the simple form \eqref{TvpVerticalProfile} used here (Online Resource Fig.~\ref{Fig8_supp}b). In practice, we removed from the analysis all of those leading to the unphysical case $\Delta T_{\textup{vp}}<0$. We have noticed that these `ill-processed' profiles dominantly occur in winter and are evenly spread across the hours of the day. Importantly, they represent $\simeq12$\% of the data only (Online Resource Fig.~\ref{Fig8_supp}c,d), and we are thus confident that this data treatment does not affect our conclusions.

\begin{figure}[p]
\centering
\includegraphics[scale=1]{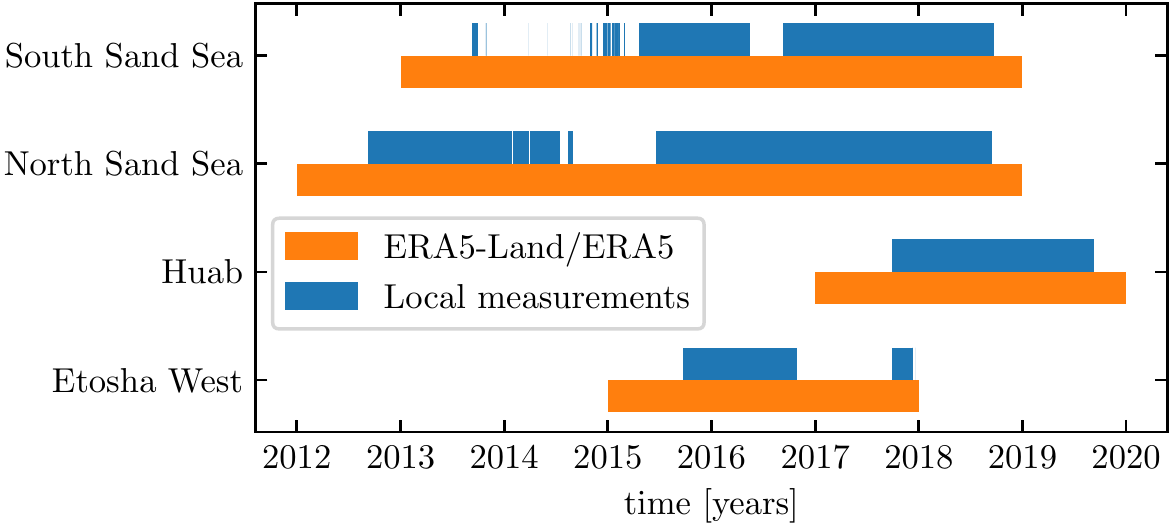}
\caption{Gantt chart representing the valid time steps for the two data sets, for all stations.}
\label{Fig1_supp}
\end{figure}

\begin{figure}[p]
\centering
\includegraphics[scale=1]{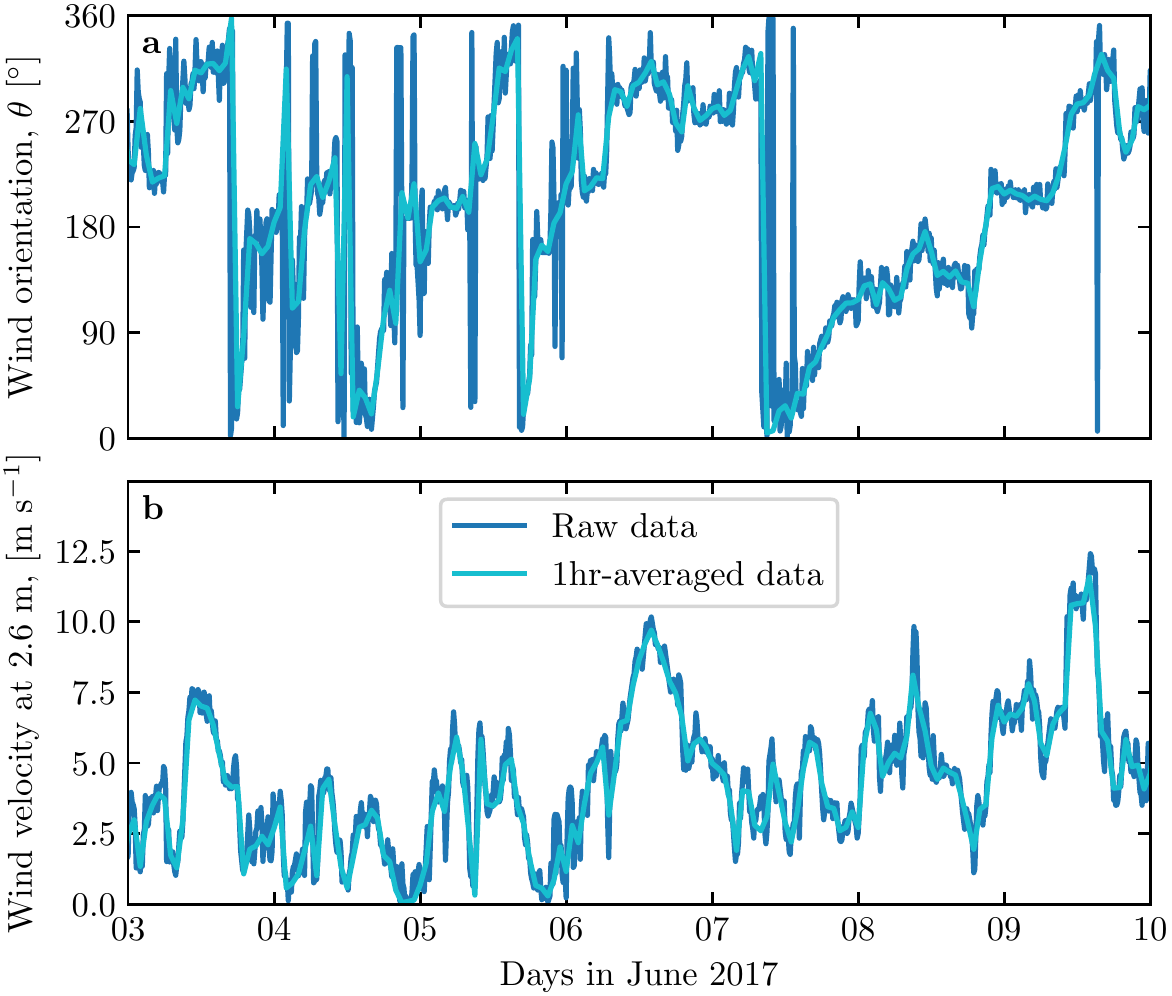}
\caption{Local wind measurements: comparison between raw (blue) and hourly-averaged (light blue) data from South Sand Sea station. \textbf{a}: wind direction. \textbf{b}: wind velocity at height $2.6$~m.}
\label{Fig2_supp}
\end{figure}

\begin{figure}[p]
\centering
\includegraphics[scale=1]{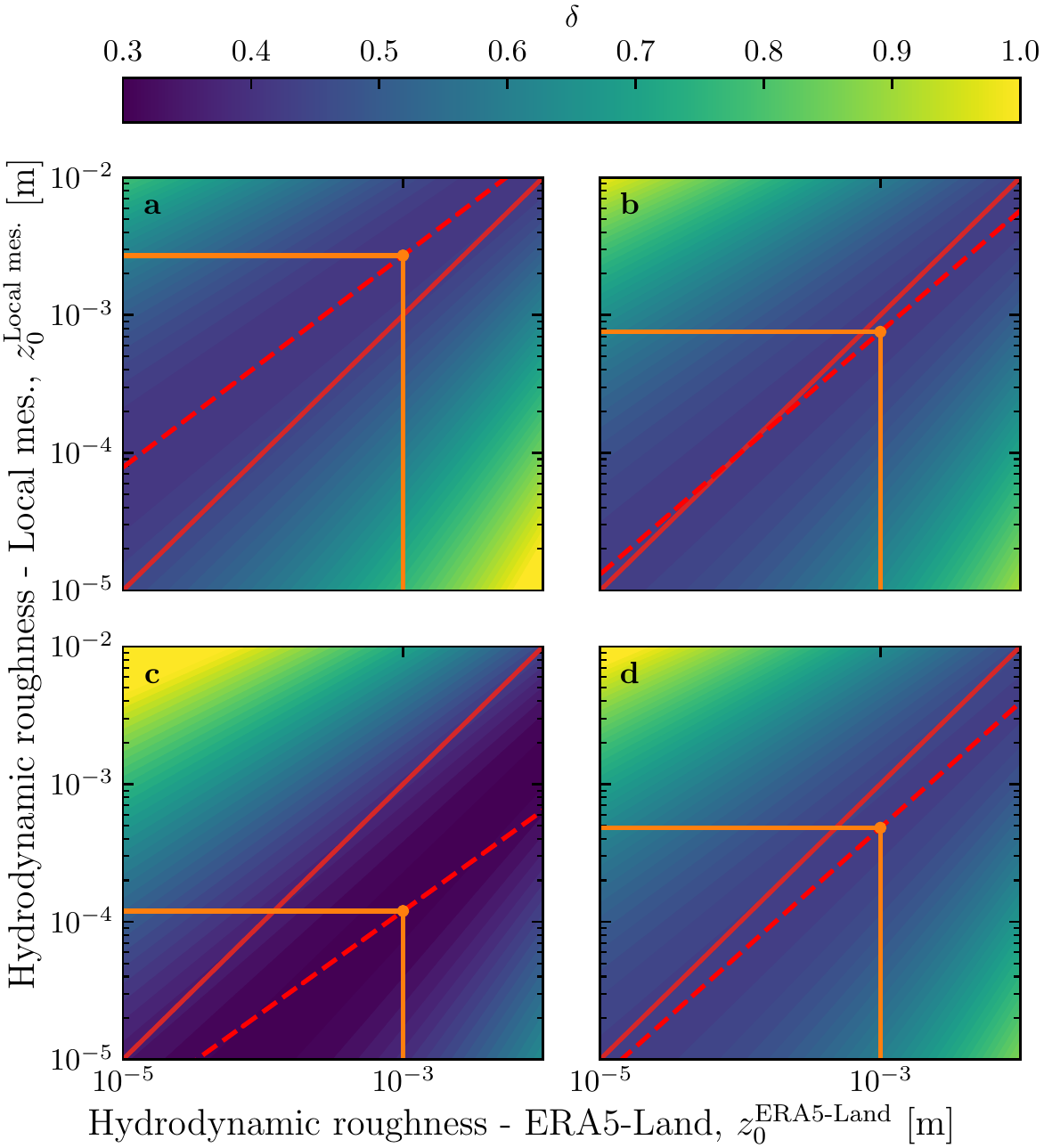}
\caption{Calibration of hydrodynamic roughness. The parameter $\delta$ \eqref{metric_roughness} quantifying the difference between local and predicted winds is shown in color scale as a function of the hydrodynamic roughnesses chosen for the ERA5-Land and for local winds, for the (\textbf{a}) Etosha West, (\textbf{b}) North Sand Sea, (\textbf{c}) Huab and (\textbf{d}) South Sand Sea stations. The red dashed and plain lines shows the minima of $\delta$ and the identity line, respectively. The orange lines and dots highlight the chosen hydrodynamic roughnesses for the local winds deduced from setting $z_{0}^{\textup{ERA5-Land}} = 1$~mm.}
\label{Fig3_supp}
\end{figure}

\begin{figure}[p]
  \centering
  \includegraphics[scale=1]{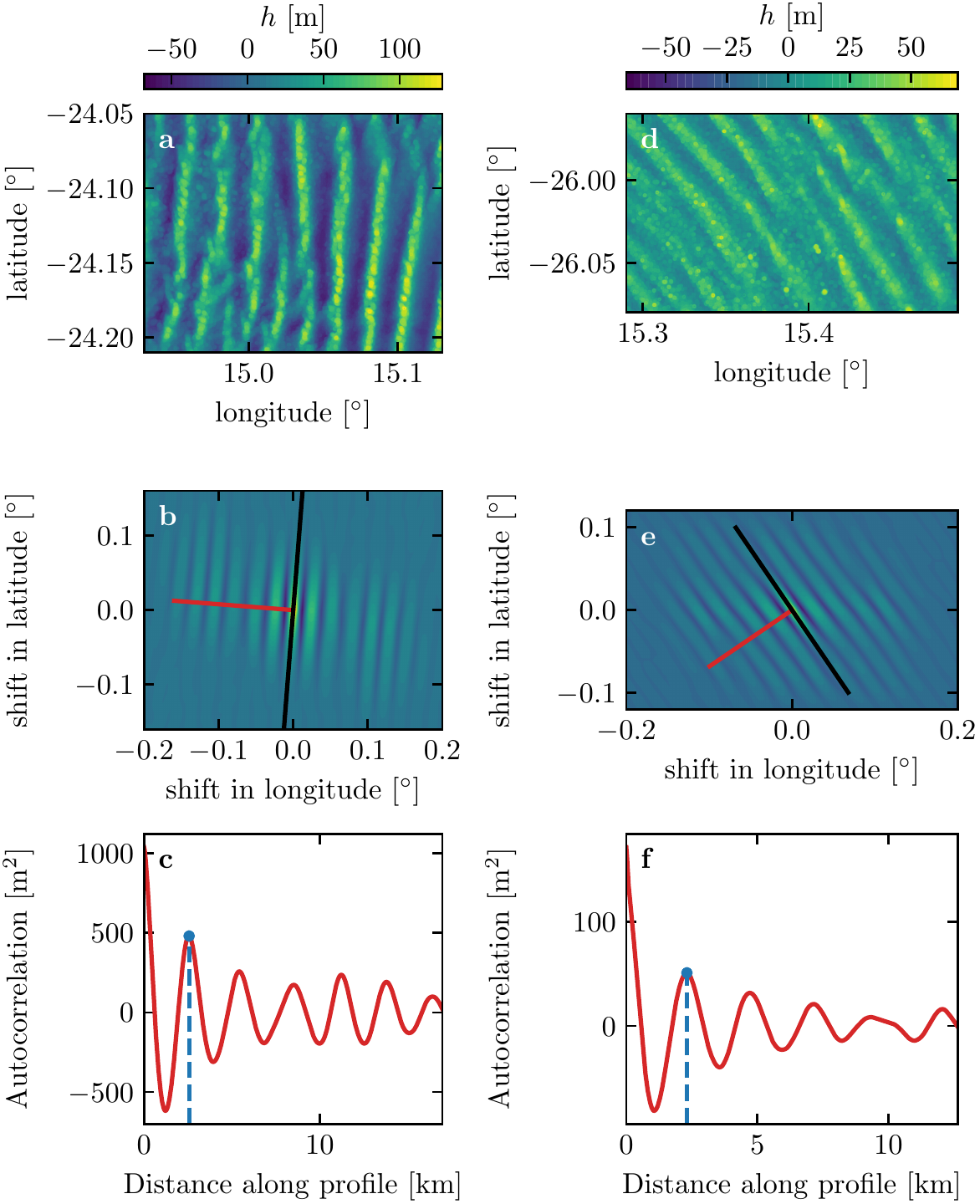}
  \caption{Analysis of the DEMs of the North Sand Sea (left column -- panels \textbf{a}, \textbf{b}, \textbf{c}) and South Sand Sea (right column -- panels \textbf{d}, \textbf{e}, \textbf{f}) stations. \textbf{a--d}: Bed elevation detrended by a fitted second order polynomial base-line. \textbf{b--e}: Autocorrelation matrix shown in color scale. The black line shows the detected dune orientation, and the red line represents the autocorrelation profile along which the dune wavelength is calculated, displayed in \textbf{c--f}. The blue lines and dots show the first peak of the autocorrelation profile, whose abscissa gives the characteristic wavelength of the dune pattern.}
  \label{Fig4_supp}
\end{figure}

\begin{figure}[p]
\centering
\includegraphics[scale=1]{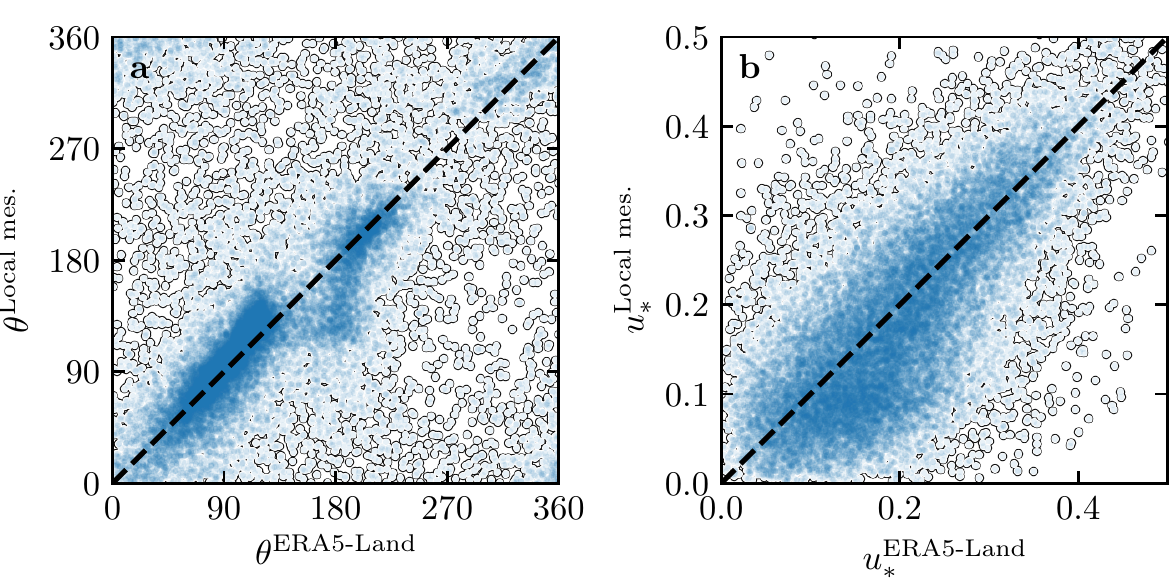}
\caption{Statistical comparison of the wind orientation (\textbf{a}) and velocity (\textbf{b}) between the ERA5-Land dataset and the local measurements for the Huab and Etosha West stations. Data point clustering around identity lines (dashed and black) provide evidence for agreement of the two sets.}
\label{Fig5_supp}
\end{figure}

\begin{figure}[p]
\centering
\includegraphics[scale=1]{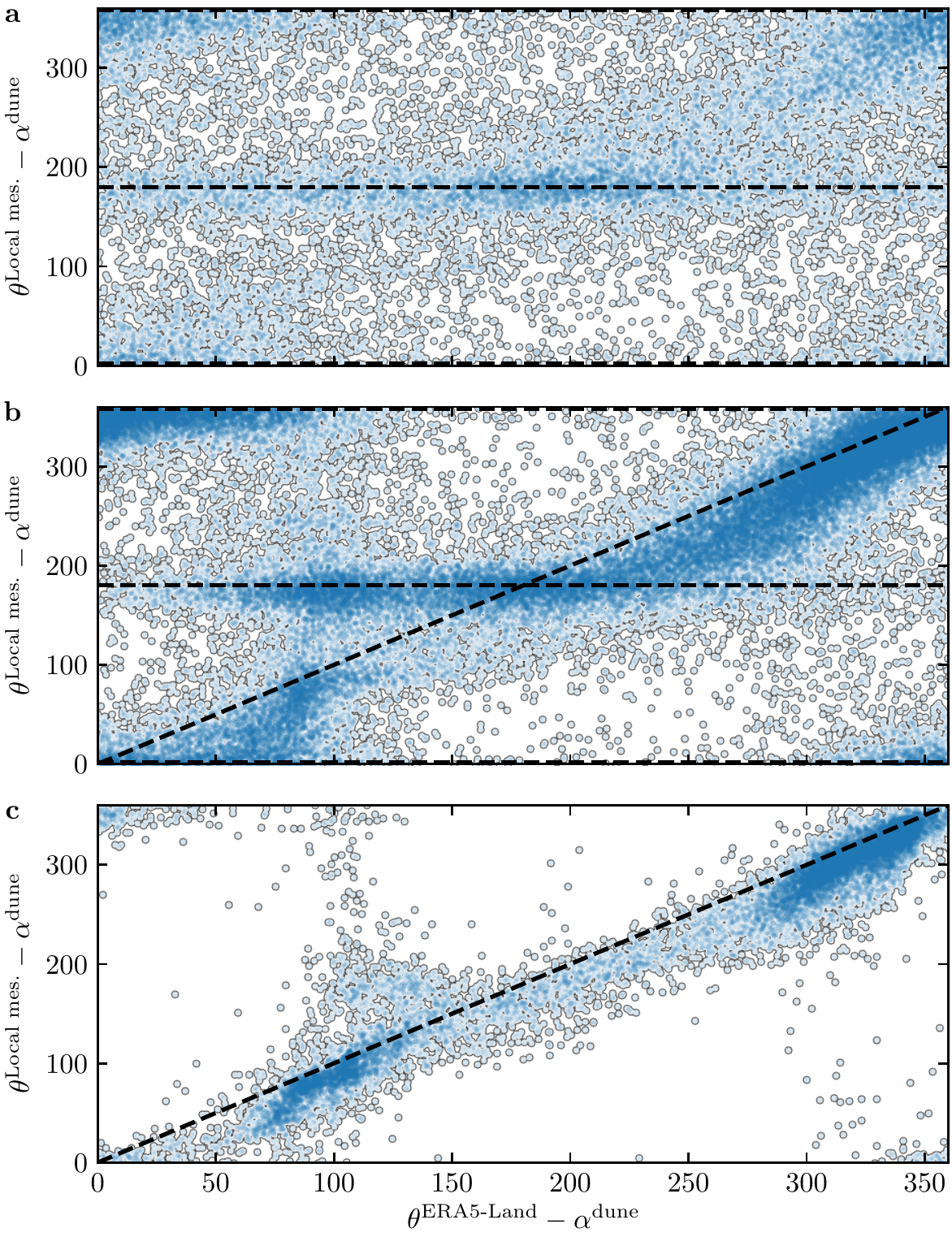}
\caption{Statistical comparison of the wind orientation between the ERA5-Land dataset and the local measurements for the South Sand Sea and North Sand Sea stations, for different velocity ranges. \textbf{a}: $u_{*}^{\textup{ERA5-Land}} < 0.1~\textup{m}~\textup{s}^{-1}$. \textbf{b}: $0.1 < u_{*}^{\textup{ERA5-Land}} \leq 0.25~\textup{m}~\textup{s}^{-1}$. \textbf{c}: $u_{*}^{\textup{ERA5-Land}} \geq 0.25~\textup{m}~\textup{s}^{-1}$. The measured dune orientations are subtracted to the wind orientation, which allows us to plot both stations on the same graph. Black dashed lines indicate locally measured orientations aligned with the dune crests (here $0^\circ$, $180^\circ$ and $360^\circ$ -- panels \textbf{a}, \textbf{b}), as well as the identity lines (panels \textbf{b}, \textbf{c}).}
\label{Fig6_supp}
\end{figure}

\begin{figure}[p]
\centering
\includegraphics[scale=1]{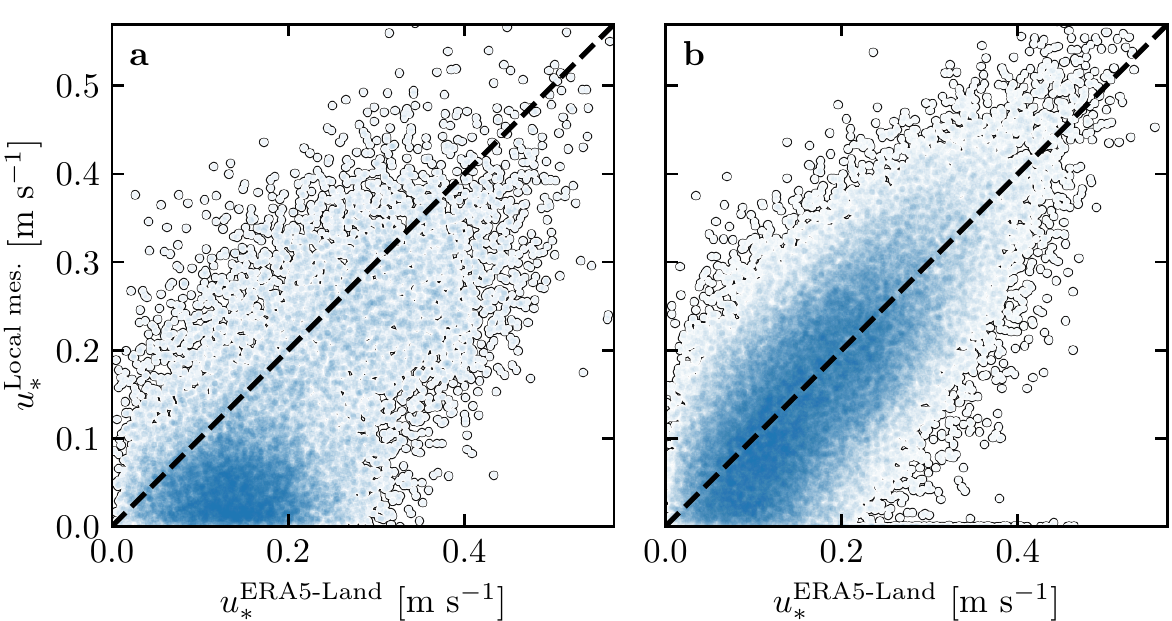}
\caption{Statistical comparison of the wind velocity between the ERA5-Land dataset and the local measurements for the South Sand Sea and North Sand Sea stations. \textbf{a}: Nocturnal summer easterly wind. \textbf{b}: Diurnal southerly wind. Black dashed lines are identity lines. The angle ranges used to select diurnal and nocturnal summer winds are the same as those in Figs.~\ref{Fig4} and Figs.~\ref{Fig6} of the main article.}
\label{Fig7_supp}
\end{figure}

\begin{figure}[p]
\centering
\includegraphics[scale=1]{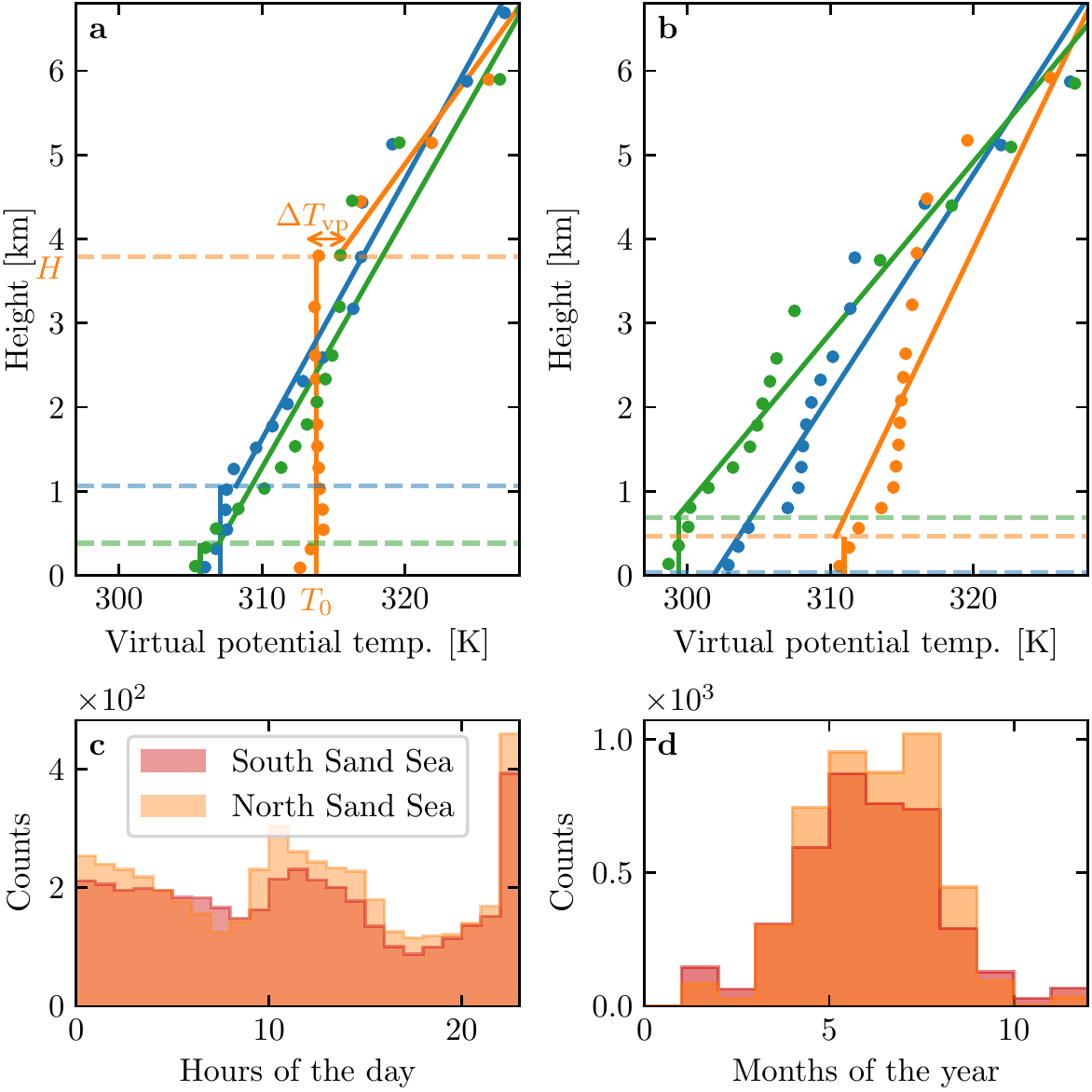}
\caption{\textbf{a}: Vertical profiles of the virtual potential temperature at three different times (blue: 29/11/2012 - 11.00 UTC, orange: 21/03/2017 - 12.00 UTC, green: 21/03/2017 - 20.00 UTC) at the South Sand Sea station. Dots: data from the ERA5 reanalysis. Dashed lines: boundary layer height given by the ERA5 reanalysis. Plain lines: vertical (ABL) and linear (FA) fits to estimate the quantities displayed in Online Resource Fig.~\ref{Fig9_supp}. \textbf{b}: Examples of ill-processed vertical profiles at three different times (blue: 2/12/2013 - 23.00 UTC, orange: 20/03/2017 - 00.00 UTC, green: 14/07/2017 - 14.00 UTC) at the South Sand Sea station. Distribution of ill-processed vertical profiles at South (orange) and North (light orange) Sand Sea station: hourly (\textbf{c}) and monthly (\textbf{d}) counts.}
\label{Fig8_supp}
\end{figure}

\begin{figure}[p]
\centering
\includegraphics[scale=1]{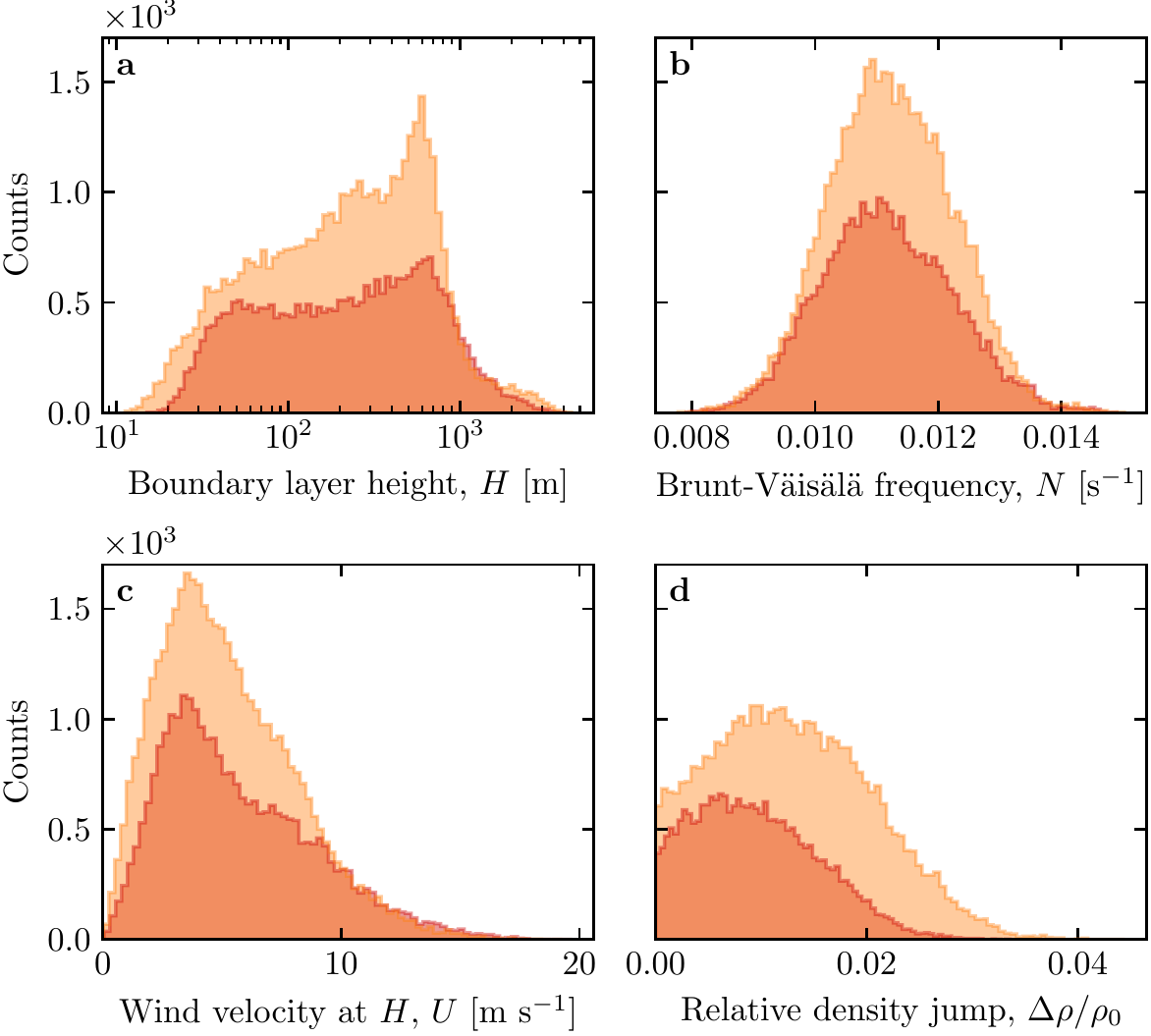}
\caption{Distributions of the meteorological parameters resulting from the processing of the ERA5-Land data for the South Sand Sea (orange) and the North Sand Sea (light orange) stations.}
\label{Fig9_supp}
\end{figure}

\begin{figure}[p]
\centering
\includegraphics[scale=1]{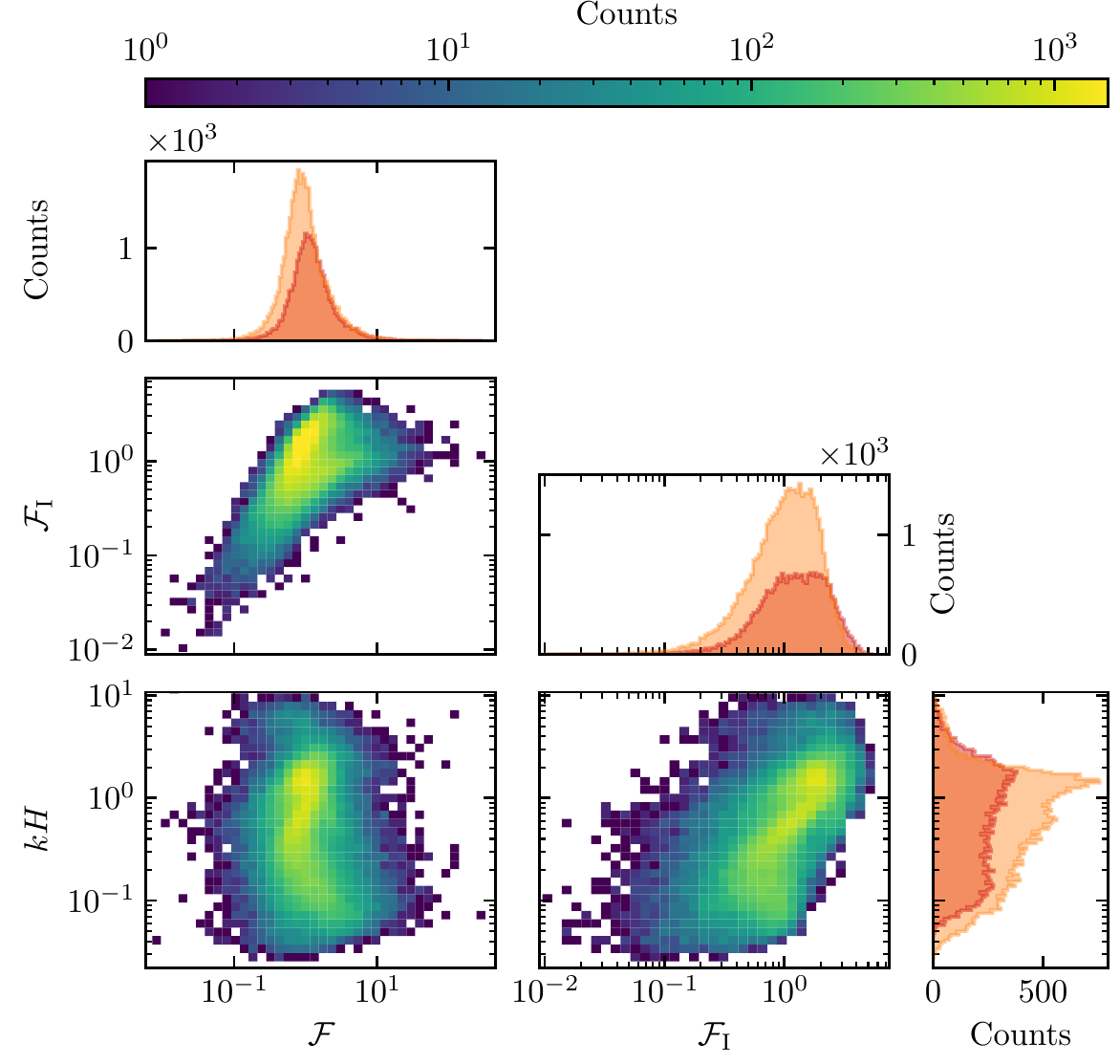}
\caption{Non-dimensional parameters distributions. For the marginal distributions, the light orange corresponds to the South Sand Sea station, and the orange to the North Sand Sea station.}
\label{Fig10_supp}
\end{figure}

\begin{figure}[p]
\centering
\includegraphics[scale=1]{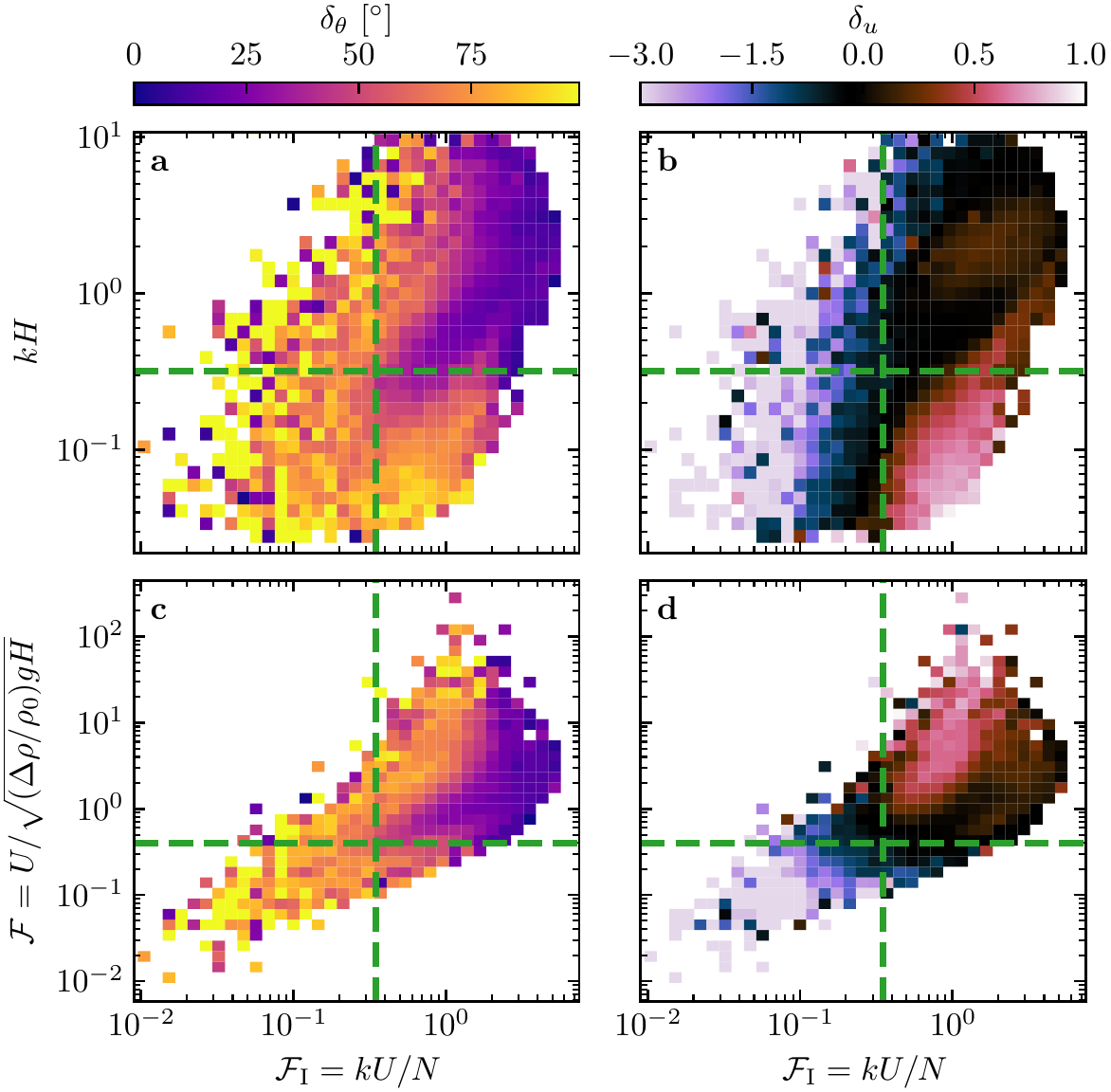}
\caption{Regime diagrams of the wind deviation $\delta_{\theta}$ and relative attenuation/amplification $\delta_{u}$ in the spaces $(\mathcal{F}_{\textup{I}}, \, kH)$ and $(\mathcal{F}_{\textup{I}}, \, \mathcal{F})$, containing the data from both the North Sand Sea and South Sand Sea stations. Green dashed lines empirically delimit the different regimes. The point density in each bin of the diagrams is shown in Online Resource Fig.~\ref{Fig10_supp} -- 95\% of the data occur in the range $-1 < \delta u < 1$. The similar regime diagrams in the space $(\mathcal{F}, \, kH)$ are shown in Fig.~\ref{Fig8}.}
\label{Fig11_supp}
\end{figure}

\begin{figure}[p]
\centering
\includegraphics[scale=1]{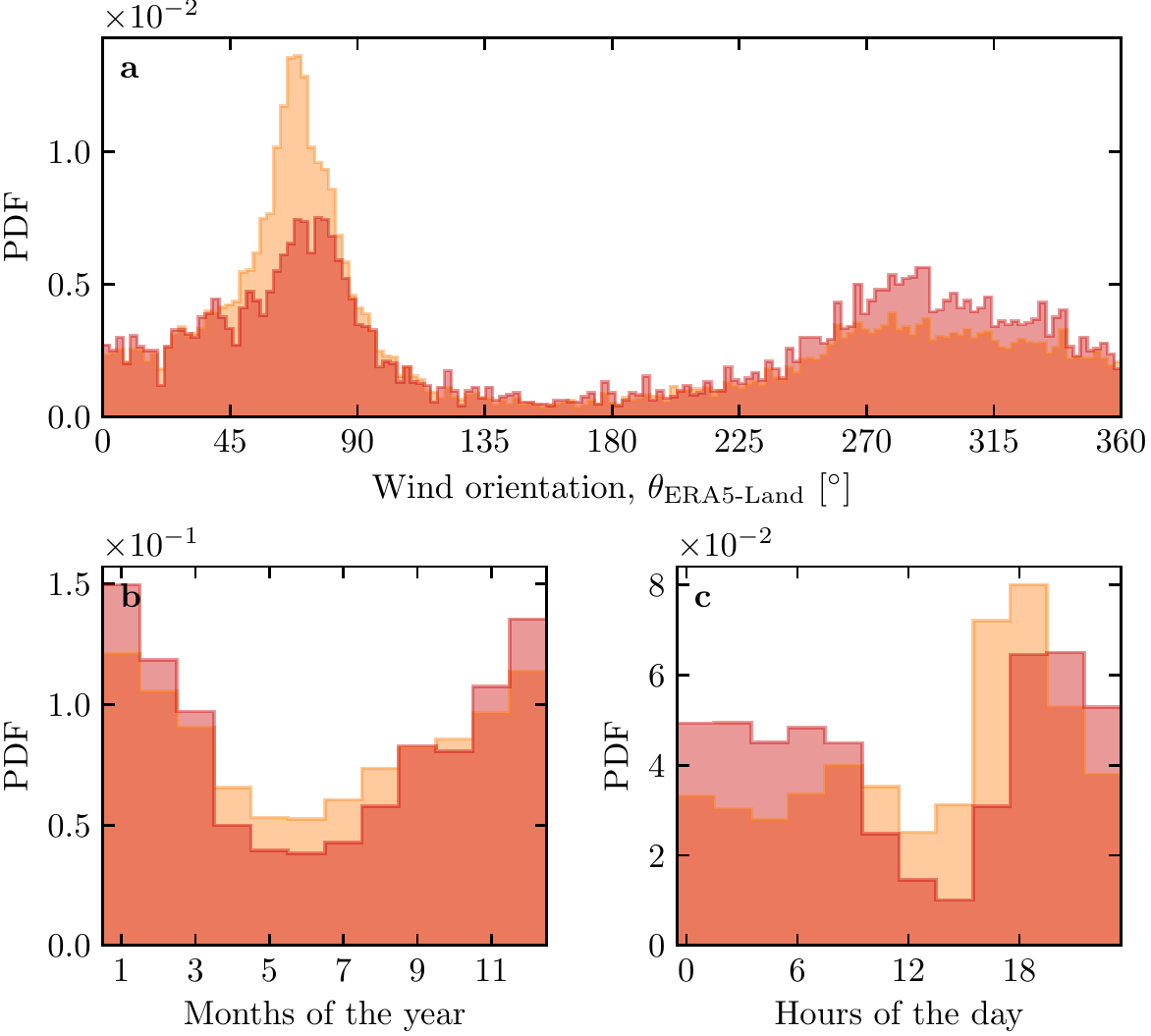}
\caption{Normalized distributions of amplified velocities for the North sand Sea (light orange: $\delta_u < 0$, orange: $\delta_u < -0.5$). \textbf{a}: Angular distributions. \textbf{b}: Monthly distributions. \textbf{c}: Hourly distributions.}
\label{Fig12_supp}
\end{figure}

\begin{figure}[p]
\centering
\includegraphics[scale=1]{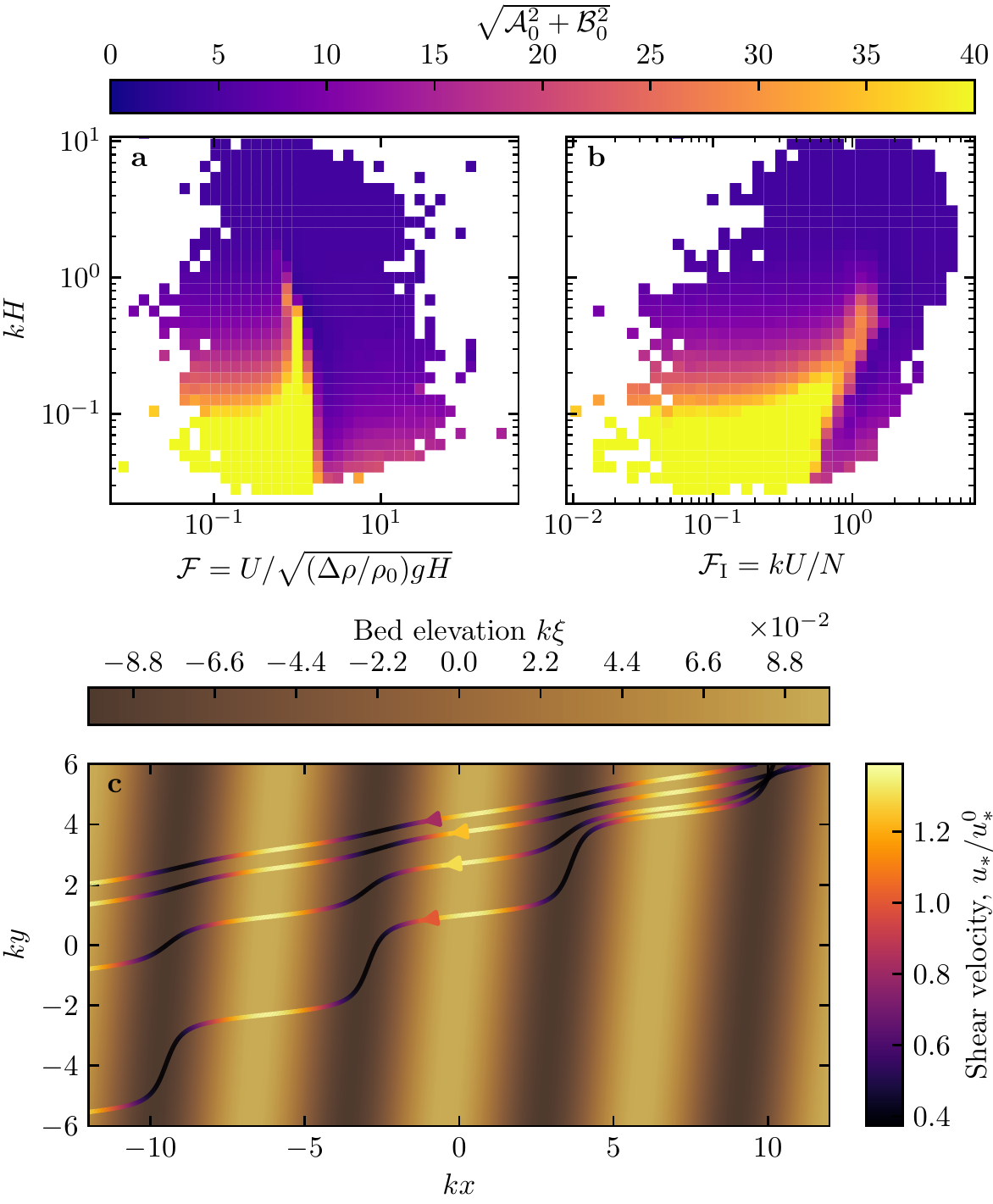}
\caption{Computation of the flow disturbance with the linear model of \citet{Andreotti2009}. \textbf{a}--\textbf{b}: Magnitude of the hydrodynamic coefficients $\mathcal{A}_0$ and $\mathcal{B}_0$, calculated from the time series of the non-dimensional numbers corresponding to the ERA5-Land wind data and ERA5 data on vertical pressure levels. \textbf{c} Shear velocity streamlines over sinusoidal ridges of amplitude $k\xi_0 = 0.1$ and for increasing values of $\sqrt{\mathcal{A}_{0}^{2} + \mathcal{B}_{0}^{2}}$. From the upper to the lower streamline, values of $\left(kH,\, \mathcal{F},\, \mathcal{F}_{\rm I},\, \mathcal{A}_{0},\, \mathcal{B}_{0},\, \sqrt{\mathcal{A}_{0}^{2} + \mathcal{B}_{0}^{2}}\right)$ are (1.9, 0.6, 1.5, 3.4, 1.0, 3.5), (1.5, 0.3, 0.4, 4.8, 1.4, 5.0), (0.1, 3.5, 1.0, 8.6, 0.1, 8.6), (0.5, 0.05, 0.04, 9.6, 2.5, 9.9).}
\label{Fig13_supp}
\end{figure}

\begin{figure}[p]
\centering
\includegraphics[scale=1]{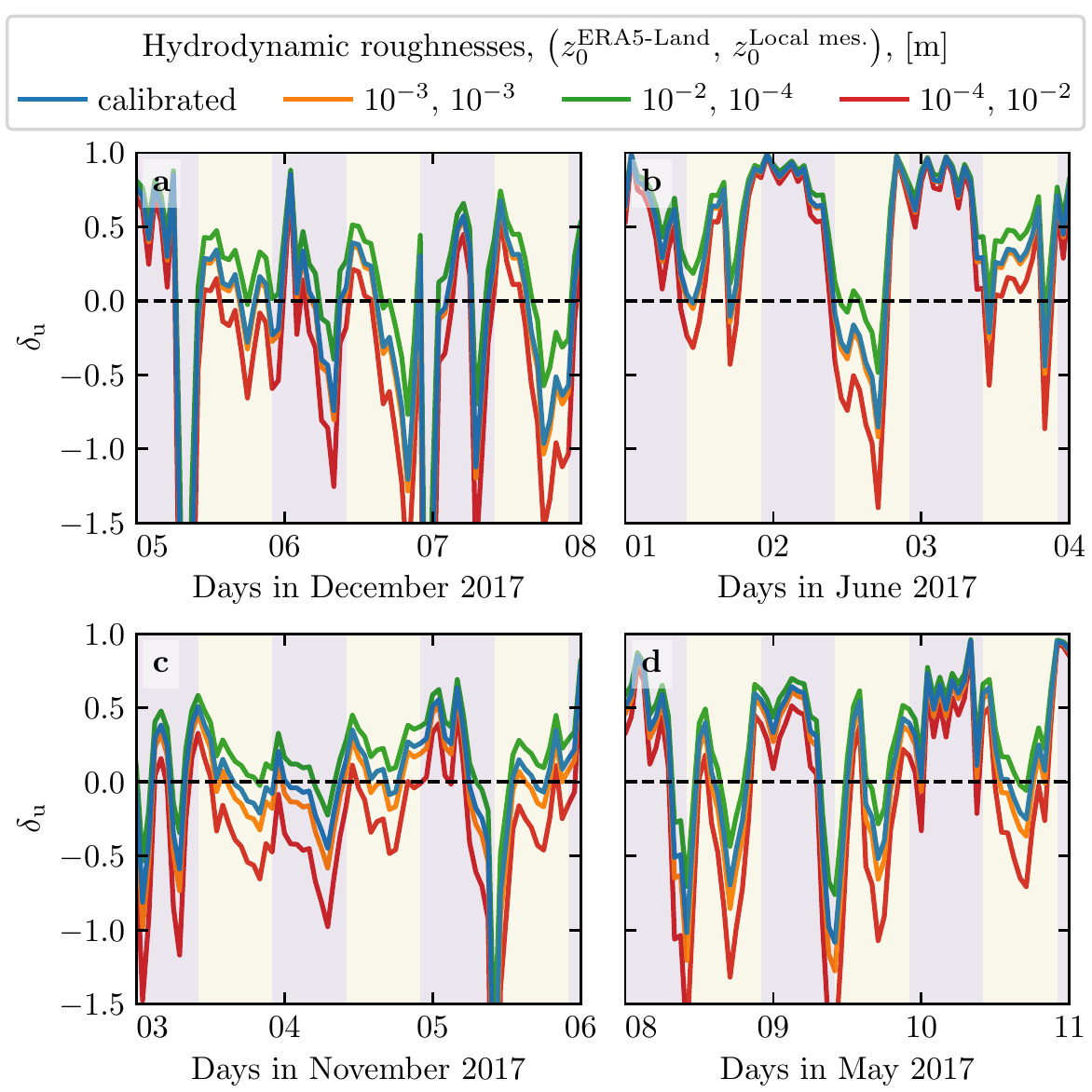}
\caption{Time series of the relative velocity disturbance $\delta_{u}$ corresponding to Fig.~\ref{Fig5}, for different values of the hydrodynamic roughnesses. \textbf{a}: North Sand Sea -- summer, \textbf{b}: North Sand Sea -- winter, \textbf{d}: South Sand Sea -- summer, \textbf{e}: South Sand Sea -- winter. Note that $\delta_{\theta}$ is independent of the choice of $z_{0}^{\textup{ERA5-Land}}$ and $z_{0}^{\textup{Local mes.}}$.}
\label{Fig14_supp}
\end{figure}

\end{document}